\documentclass[review]{elsarticle}
\usepackage{amsmath,amssymb,amsfonts}
\usepackage{algorithmic}
\usepackage{graphicx}
\usepackage{textcomp}
\usepackage{subfigure}
\usepackage{booktabs} 
\usepackage{lineno,hyperref}
\usepackage{xcolor}

\modulolinenumbers[5]
\newcommand{\figsize}{0.625}
\usepackage{multirow}
\DeclareMathOperator*{\argmin}{argmin}

\journal{Journal of \LaTeX\ Templates}
\newtheorem{problem}{Finite Horizon Optimal Control Problem}
\newtheorem{remark}{Remark}

\begin{document}

\begin{frontmatter}
\title{Optimal Charging of an Electric Vehicle Battery Pack: A Real-Time Sensitivity-Based MPC approach}

%\author{Andrea~Pozzi, Marcello~Torchio, Richard~Braatz, Davide M. Raimondo% <-this % stops a space
%\thanks{A. Pozzi and D. M. Raimondo are with the Department
%of Electrical and Computer Engineering, University of Pavia, Pavia,
% 27100 Italy e-mail: andrea.pozzi03@universitadipavia.it, davide.raimondo@unipv.it.}% <-this % stops a space
%\thanks{M. Torchio is with the Department
%of Civil Engineering, University of Pavia, Pavia,
% 27100 Italy e-mail: marcello.torchio01@universitadipavia.it.}
% \thanks{R. Braatz is with the Massachusetts Institute of Technology,
%Cambridge, MA, 02139 e-mail: braatz@mit.edu.}
%\thanks{Manuscript received May 1, 2019.}}

\author{Andrea Pozzi\corref{mycorrespondingauthor}}
\cortext[mycorrespondingauthor]{Corresponding author}
\ead{andrea.pozzi03@universitadipavia.it}
\address{University of Pavia, Via Ferrata 5, 27100, Pavia}
\author{Marcello Torchio}
\address{United Technologies Research Centre Ireland Ltd, 2nd Floor Penrose Business Centre, Cork, Ireland}
\author{Richard D. Braatz}
\address{Massachusetts Institute of Technology, 77 Massachusetts Avenue, Cambridge, MA, 02139}
\author{Davide M. Raimondo}
\address{University of Pavia, Via Ferrata 5, 27100, Pavia}

%
%%%% or include affiliations in footnotes:

\begin{abstract}
Lithium-ion battery packs are usually composed of hundreds of cells arranged in series and parallel connections. The proper functioning of these complex devices requires suitable Battery Management Systems (BMSs). Advanced BMSs rely on mathematical models to assure safety and high performance. While many approaches have been proposed for the management of single cells, the control of multiple cells has been less investigated and usually relies on simplified models such as equivalent circuit models. This paper addresses the management of a battery pack in which  each cell is explicitly modelled as the Single Particle Model with electrolyte and thermal dynamics. A nonlinear Model Predictive Control (MPC) is presented for optimally charging the battery pack while taking voltage and temperature limits on each cell into account. Since the computational cost of nonlinear MPC grows significantly with the complexity of the underlying model, a sensitivity-based MPC (sMPC) is proposed, in which the model adopted is obtained by linearizing the dynamics along a nominal trajectory that is updated over time.
The resulting sMPC optimizations are quadratic programs which can be solved in real-time even for large battery packs (e.g. fully electric motorbike with $156$ cells) while achieving the same performance of the nonlinear MPC. 

\end{abstract}

\end{frontmatter}

%\linenumbers

\section{Introduction}
\label{sec:Introduction}
Electric vehicle battery packs usually consist of several cells which are arranged in series and parallel connections in order to meet power and capacity requirements. The proper functioning of these complex devices requires Battery Management Systems (BMSs) \cite{Lu2013}. One of the main tasks of a BMS is to safely charge the battery. This objective is usually addressed through standard protocols, such as Constant Current or Constant Current--Constant Voltage  \cite{Notten2005,Keil2016}. These methods rely on fixed voltage limits that assure reasonable performance for each cell composing the pack over its lifetime.
In practice, these approaches lead to a suboptimal exploitation of the battery, in which some cells are undercharged and some are overcharged. As a consequence, some cells may experience fast degradation, thermal runaway, and, in certain cases, even explosions.  
These problems can be significantly alleviated if advanced BMSs, which rely on mathematical models of the pack, are employed \cite{Chaturvedi2010}.

In this work, we focus on batteries based on lithium-ion chemistries which, thanks to their unique characteristics, have proven to be the most promising energy accumulators for many applications \cite{Dunn2011, Tarascon2011}. Within this context, the two main categories of cell models employed in advanced BMSs are: Equivalent Circuit Models (ECMs, \cite{Hu2012}) and Electrochemical Models (EMs, \cite{Gomadam2002,Santhanagopalan2006}). While the former are simple and intuitive, the latter provide a detailed description of the electrochemical phenomena which occur inside a cell. Among the EMs, the Pseudo-Two-Dimensional (P2D) model \cite{Doyle1993} -- also known as Doyle-Fuller-Newman -- is the most widely used. This model consists of coupled and nonlinear Partial Differential Algebraic Equations (PDAEs).  Due to its high computational cost, the P2D is more suited for simulation purposes rather than for on-board control applications. Moreover, the use of the P2D model within a control framework is limited by identifiability and observability issues \cite{LopezC2016,Moura2015}.  
For all these reasons, the research community has been interested in the development of simplified electrochemical models, which are faster to simulate, are identifiable and observable, and still provide a reasonable description of the internal cell phenomena \cite{Zou2014}. Among them, the Single Particle Model (SPM) \cite{Ning2004,Santhanagopalan2006}), which is derived from the P2D by modelling the electrodes as single particles, has received a lot of attention. The parameters in the SPM are fit to input-output data from the battery. The parameter identifiability and state observability of the SPM have been analyzed in several works, e.g. \cite{DiDomenico2010,Bizeray2018,Pozzi2018c}. While the SPM has much lower computational cost than the P2D, which is an advantage for control purposes, the SPM neglects electrolyte, thermal, and ageing dynamics.
The model fidelity has been increased by extending the SPM to include electrolyte dynamics (SPMe) \cite{Moura2017} and thermal dynamics (SPMeT) \cite{Perez2016}.

The model-based charging of lithium-ion batteries has been addressed by many authors over the years. Most of the literature focuses on the control of a single lithium-ion cell. Within this context, different control approaches have been considered, such as fuzzy logic \cite{Hsieh2001, Huang2009, Asadi2012, Wang2014}, empirical rules \cite{Purushothaman2006,Chen2008,Chen2008a,Khan2018}, and optimization-based strategies \cite{Liu2009,Klein2011,Moura2013, Xavier2015,Torchio2015,Perez2016,
Lucia2017,Romagnoli2017}. 
%Zhang2017, Liu2018 (optimal-rule based trovati da ray)
Model Predictive Control (MPC) \cite{Maciejowski2002,Camacho2013} appears to be the most used optimization-based methodology for charging of cells \cite{Yan2011,Klein2011,Xavier2015,Torchio2015,Torchio2016b,Torchio2017,
Lucia2017,Zou2017, Zou2018, Pozzi2018a,Pozzi2018b}. MPC is particularly appropriate for controlling multivariable nonlinear systems while taking an objective function and constraints on both inputs and states into account. In the context of lithium-ion cells, MPC aims to minimize the charging time while satisfying temperature and voltage constraints.
%The authors in \cite{Romagnoli2017} have proposed a reference governor charging technique based on an equivalent hydraulic model of a single cell
In particular, the works in \cite{Yan2011, Xavier2015, Pozzi2018b} have proposed  MPC strategies based on ECMs, while \cite{Klein2011, Lucia2017,Pozzi2018a, Zou2018} have suggested the use of electrochemical models together with MPC in order to achieve better performance. In order to alleviate the complexity of physics-based models, input-output descriptions have been used in combination with MPC in \cite{Torchio2015} and \cite{Torchio2016b}. In particular,  \cite{Torchio2015} relied on a linear input-output model, while a piecewise Auto Regressive Exogenous (ARX) input-output model was proposed in \cite{Torchio2016b}. Finally, in order to achieve a reasonable tradeoff between computational time and accuracy in the prediction, the usage of Linear Time Varying (LTV) models for MPC has been addressed \cite{Torchio2017, Zou2017}. The works listed above focused on the charging of a single lithium-ion cell. 

Batteries composed of several cells are necessary in many applications, such as hybrid electric vehicles. The optimal charging of battery packs has been less investigated than the case of single cells. In particular, most of the available literature relies, as control models, on very simple lumped ECMs (see e.g. \cite{Liu2008, Moura2010}). Few works tackle the optimal control of lithium-ion batteries by directly modelling each cell individually. This level of detail is necessary in important tasks such as the model-based State of Charge (SOC) balancing of series-connected cells \cite{Einhorn2011, Danielson2012, Caspar2014,Altaf2017, McCurlie2017, Pozzi2019}. Most of the research produced in this area relies on linear ECMs for each cell (with the exception of \cite{Pozzi2019} which is based on the SPMeT). The usage of such models in the battery optimal control context has the advantage of a low computational cost, which allows for real-time implementation even in the case of a high number of cells. On the other side, simple linear models  often fail to grasp the real behaviour of the battery pack. As a consequence, the resulting charging may be suboptimal and may not even satisfy the safety constraints on all the different cells.
The use of electrochemical models within a control framework would be a possibility, but comes at the price of a prohibitive computational cost.
As a compromise, the use of linearized  electrochemical models seems promising in order to achieve high performance with a reasonable computational load.

In this work, we consider a battery pack composed by series-connected modules, each of which are constituted of parallel-connected cells modelled according to the SPMeT. The resulting model consists of a set of nonlinear Differential Algebraic Equations (DAEs), for which suitable linearization methods are applied. In the following, in order to provide high accuracy for the model used for the  control, we rely on a sensitivity-based linearization method, which has been described in detail in \cite{DeOliveira1995, Santin2016} for a system of Ordinary Differential Equations (ODE) and here extended to the case of a system of DAEs. Such an approach differs from standard LTV techniques \cite{Campbell1995}, since the sensitivities of the states and outputs to input variations are continuously integrated together with the model equations rather than evaluated only at discrete time steps.

The main contribution of this work is the application of a sensitivity-based linear MPC  to the management of large battery packs, taking into account safety constraints (such as temperature and voltage limits) on all the different cells. To the best of our knowledge, this is the first time that such an approach is used in the context of the control of battery packs. As shown in the Results section, a detailed analysis is conducted by comparing a standard nonlinear MPC and the Constant-Current Constant-Voltage algorithm with the proposed sensitivity-based MPC over an increasing number of series and parallel connected cells. The sensitivity-based MPC has much lower computational cost than nonlinear MPC while achieving comparable performance. Since the choice of the nominal trajectory is fundamental in order to obtain an accurate linearization, we also provide an adaptive method in order to update the nominal trajectory during the charge. In addition, the need for an optimal management of the different cells is made evident by highlighting the disadvantages of standard charging protocols, such as the CC-CV. The value of the proposed methodology is demonstrated by application to the optimal management of an electric motorbike battery pack composed by 156 cells. In this application, the sensitivity-based MPC provides optimal performance with a computational cost compatible with the sampling time. 

The paper is organized as follows. Section \ref{sec:Model} recalls the equations of the model used in the control algorithm, while Section \ref{sec:Method} proposes the sensitivity-based  model predictive control algorithm. The adapting of this MPC formulation for the context of lithium-ion batteries and the main results are provided in Section \ref{sec:results}. Finally, Section \ref{sec:conclusions} concludes the paper.

\section{Model}
\label{sec:Model}
In this section, the main equations of the simplified electrochemical model proposed in \cite{Pozzi2019} are recalled.  The latter consist of a reduction of the P2D \cite{Doyle1993} model which is detailed but also suitable for control purposes. In particular, starting from the SPMe \cite{Moura2017}, the Partial Differential Equation (PDE) which describes the diffusion of ions within the electrolyte is spatially discretized according to the finite volume method \cite{Eymard2000}. Moreover, a polynomial approximation of the ion concentration along the radial axis of each electrode is considered in order to reduce the Fick's laws into Ordinary Differential Equations (ODEs) \cite{Subramanian2005}, as previously done in the context of SPMe by \cite{Pozzi2018}. 
In addition, the thermal dynamics are described by adapting the equations proposed in \cite{Perez2016,Perez2017}, where the heat generated by the cell is transferred to a proper cooling system. The cell model is presented in Section \ref{sub:model_of_a_single_cell}, while the equations of the battery pack model are given in Section \ref{sub:battery_pack_model}.

\subsection{Model of a Single Cell} \label{sub:model_of_a_single_cell}
In this subsection, the index $j \in \{p,s,n\}$ refers to all the cell sections, while the index $i \in \{p,n\}$ is used in equations valid only for the electrodes. The independent variables $x\in \mathbb{R}$ and $r \in \mathbb{R}$ are the axial and radial coordinates respectively, and $t \in \mathbb{R}$ is the time. 
According to the approximation in \cite{Subramanian2005}, the ion concentration along the radial axis $r$ of each electrode is described by a fourth-order polynomial function of $r$. This approach results in coefficients that are functions of the solid average concentration $\bar c_{s,i}(t)$ and the average concentration flux $\bar q_{i}(t)$. In the following, we introduce the dynamics of such variables that are necessary for reconstructing the value of the ion concentration along the radial axis. Consider the average stoichiometry in the electrodes defined by
\begin{align}
\bar \theta_i (t)=\frac{\bar c_{s,i}(t)}{c_{s,i}^{max}}
\end{align}
where $c_{s,i}^{max}$ is the maximum solid concentration. 
Relying on the fact that the moles of lithium in the solid phase are preserved \cite{DiDomenico2010}, the average stoichiometry in the anode can be expressed in terms of the cathode by
\begin{align}
\bar \theta_n(t)= \theta^{0\%}_n+\frac{\bar \theta_p(t)-\theta^{0\%}_p}{\theta_p^{100\%}-\theta^{0\%}_p} (\theta_n^{100\%}-\theta^{0\%}_n )
\end{align}
where $\theta_{i}^{0\%}$ and $\theta_{i}^{100\%}$ represent the value of the stoichiometries  respectively when the cell is fully discharged and completely charged. 
From \cite{Subramanian2005}, the temporal evolution of the average stoichiometry in the positive electrode can be approximated by
\begin{align}
\dot{\bar \theta}_p(t)=\frac{3 I_{app}(t)}{a_p R_{p,p} L_p F A }
\end{align}
where  $I_{app}(t)$ is the applied current input (defined as being negative during charging by convention), $R_{p,i}$ is the particle radius, $L_j$ is the thickness of the $j$th section, $F$ is the Faraday constant, $A$ is the contact area between solid and electrolyte phase, and $a_i=\frac{3 \epsilon_i^{act}}{R_{p,i}}$ is the specific active surface area, with the active material volume fraction $\epsilon_i^{act}$ defined by
\begin{subequations}
\begin{align}
\epsilon^{act}_p&=-\frac{C}{\Delta \theta_p A F L_p c_{s,p}^{max}}\\
\epsilon^{act}_n&=\frac{C}{\Delta \theta_n A F L_n c_{s,n}^{max}}
\end{align}
\end{subequations}
in which  $C$ is the cell capacity and $\Delta \theta_i= \theta_i^{100\%}- \theta_i^{0\%}$.
In accordance with \cite{Subramanian2005}, the volume-averaged concentration fluxes can be described by 
\begin{subequations}\label{eq:average_concentration_flux}
\begin{align}
	\dot {\bar q}_p(t) &= -30 \frac{D_{s,p}(T(t))}{R_{p,p}^2}\bar q_p(t) + \frac{45}{2R_{p,p}^2 F A L_p a_p}I_{app}(t)\\
		\dot {\bar q}_n(t) &= -30 \frac{D_{s,n}(T(t))}{R_{p,n}^2}\bar q_n(t) - \frac{45}{2R_{p,n}^2 F A L_n a_n}I_{app}(t)
\end{align} 
\end{subequations}
where 
\begin{align}\label{eq:arrenhius}
D_{s,i}(T(t))=D_{s,i}^0 e^{\frac{-E_{a,D_{s,i}}}{RT(t)}}
\end{align}
is the solid diffusion coefficient for the $i$th section which depends on the cell temperature $T(t)$ according to the Arrenhius law,
$D_{s,i}^0$ is a pre-exponential coefficient which is assumed to be constant, $E_{a,D_{s,i}}$ is the activation energy associated with the parameter $D_{s,i}(T(t))$, and $R$ is the universal gas constant. 
Then, in accordance with \cite{Subramanian2005}, the surface stoichiometries in the positive and negative electrodes are respectively given by the algebraic equations
\begin{subequations}\label{eq:stoichiometric_surface}
\begin{align}
\theta_p(t)&= \bar \theta_p(t)+\frac{8R_{p,p}\bar q_p(t)}{35c_{s,p}^{max}}+\frac{R_{p,p} I_{app}(t)}{35D_{s,p}(T(t))F A L_p a_p c_{s,p}^{max}}\\
\theta_n(t)&= \bar \theta_n(t)+\frac{8R_{p,n}\bar q_n(t)}{35c_{s,n}^{max}}-\frac{R_{p,n} I_{app}(t)}{35D_{s,n}(T(t))F A L_n a_n c_{s,n}^{max}}  
\end{align}
\end{subequations}
Finally, the  state of charge  $SOC(t)$ is defined as
\begin{align}\label{eq:soc}
SOC(t)=100\frac{\bar \theta_{n}(t)-\theta_n^{0\%}}{\theta_n^{100\%}-\theta_n^{0\%}}
\end{align}

In this work, the PDEs governing the diffusion of the electrolyte concentration $c_{e,j}(x,t)$  \cite{Moura2017} are discretized according to the FV method, as previously done in this context by the authors in \cite{Torchio2016}, where the spatial domain is divided into $P$ non-overlapping volumes for each section. The $k$th volume, with $k=1, \cdots, P$, of the $j$th section is centered at the spatial coordinate $x_{j,k}$ and spans the interval $\Omega_{j,k}=\left[x_{j,\bar{k}},x_{j,\underline{k}}\right]$, whose  width is  $\Delta x_j=L_j/P$.   Defining $c_{e,j}^{[k]}(t)$ as the average electrolyte concentration over the $k$th volume of $j$th section gives 
\begin{subequations}
\label{eq:ce}
\begin{align}
\epsilon_p \frac{\partial  c^{[k]}_{e,p}(t)}{\partial t}=&  \left.\left[\frac{\tilde{D}_e(x,T(t))}{\Delta x_p}   \frac{\partial c_{e,p}(x,t)}{\partial x} \right]\right|_{x_{p,\underline{k}}}^{x_{p,\bar{k}}}-\frac{1-t_+}{FAL_p}I_{app}(t)\\
\epsilon_s \frac{\partial  c^{[k]}_{e,s}(t)}{\partial t}=&  \left.\left[\frac{\tilde{D}_e(x,T(t))}{\Delta x_s}   \frac{\partial c_{e,s}(x,t)}{\partial x} \right]\right|_{x_{s,\underline{k}}}^{x_{s,\bar{k}}}\\
\epsilon_n \frac{\partial  c^{[k]}_{e,n}(t)}{\partial t}=& \left.\left[\frac{\tilde{D}_e(x,T(t))}{\Delta x_n}   \frac{\partial c_{e,n}(x,t)}{\partial x} \right]\right|_{x_{n,\underline{k}}}^{x_{n,\bar{k}}} +\frac{1-t_+}{FAL_n}I_{app}(t)
\end{align}
\end{subequations}
where $t_+$ is the transference number, $\epsilon_j$ is the material porosity, and the other terms are evaluated as explained in detail in \cite{Torchio2016}. In particular, the electrolyte diffusion coefficients $\tilde{D}_e(x,T(t))$ are computed from
\begin{align}
\tilde{D}_e(x,T(t))=
\begin{cases} D_{e,1} &\mathrm{if} \quad x \in \{x_{p,\underline{P}},x_{s,\overline{1}}\}\\
D_{e,2} \quad  &\mathrm{if} \quad x \in \{x_{s,\underline{P}},x_{n,\overline{1}}\}\\
D^{eff}_{e,j}(T(t)) &\mathrm{otherwise}
\end{cases}
\end{align}
with
\begin{subequations}
\begin{align}
D_{e,1}&=\mathcal{H}\left(D^{eff}_{e,p}(T(t)),D^{eff}_{e,p}(T(t)),\Delta x_p,\Delta x_s\right)\\ 
D_{e,2}&=\mathcal{H}\left(D^{eff}_{e,s}(T(t)),D^{eff}_{e,n}(T(t)),\Delta x_s,\Delta x_n\right) 
\end{align}
\end{subequations}
where $\mathcal{H}$ is the harmonic mean operator, defined as
\begin{align}
\mathcal{H}\left(\rho_1,\rho_2,\lambda_1,\lambda_2\right)=\frac{\rho_1 \rho_2 (\lambda_1 + \lambda_2)}{\rho_1\lambda_2+\rho_2\lambda_1}
\end{align} 
and $D_{e,j}^{eff}(T(t)) = D_e(T(t))\epsilon_j^{p_j}$ with  $p_j$ being the Bruggeman coefficient and $D_e(T(t))$ being the diffusion coefficient within the electrolyte which depends on the temperature according to the Arrenhius law as in \eqref{eq:arrenhius}.

The terminal voltage is given by
\begin{align}\label{eq:voltage}
	V(t)& =  -I_{app}(t)R_{sei}+\bar U_p(t) - \bar U_n(t) + \bar \eta_p(t) - \bar \eta_n(t) +  \Delta\Phi_e(t) 
\end{align}
where $R_{sei}(t)$ is the SEI resistance, the Open Circuit Potentials (OCPs) in the positive and negative electrodes are given by
\begin{subequations}
\begin{align}\label{eq:OCPs_fun}
\begin{split}
\bar U_p(t)=&18.45\theta_p^6(t)-40.7 \theta_p^5(t)+20.94\theta_p^4(t)\\&+8.07\theta_p^3(t)-7.837\theta_p^2(t) + 0.02414\theta_p^1(t)+4.571
\end{split}
\\
\bar U_n(t)=&    \frac{0.1261\theta_n(t)+0.00694}{\theta_n^2(t)+0.6995\theta_n(t)+0.00405}
\end{align}
\end{subequations}
whose expressions in terms of surface stoichiometries are obtained by fitting the experimental data collected in \cite{Ecker2015} and depend on the considered cell (in this case, Kokam SLPB 75106100), and $\bar \eta_p(t)$ and $\bar \eta_n(t)$  are the overpotentials for the positive and negative electrodese given by
\begin{subequations}\label{eq:overpotentials}
\begin{align}
	\bar \eta_p(t) &= \frac{2RT(t)}{F}\sinh^{-1} \left(\frac{-I_{app}(t)}{2 A L_p a_p \bar i_{0,p}(t)}			\right)  \\
		\bar \eta_n(t) &= \frac{2RT(t)}{F}\sinh^{-1} \left(\frac{I_{app}(t)}{2 A L_n a_n \bar i_{0,n}(t)}			\right)
\end{align}
\end{subequations}
with the exchange current density defined as
\begin{align}
	\bar i_{0,i}(t) = Fk_i(T(t))\sqrt{\bar c_{e,i}(t)\theta_i(t)(1-\theta_i(t))}
\end{align}
where $k_i(T(t))$ is the temperature-dependent (according to the Arrenhius law) rate reaction constant and $\bar c_{e,i}(t)$ is the average electrolyte concentration in the $i$th section, approximated by
\begin{align}
\bar c_{e,i}(t)=\frac{1}{P}\sum_{k=1}^P c_{e,i}^{[k]}(t).
\end{align} 
Moreover, $\Delta\Phi_e(t)$ is computed as 
\begin{align}
	\Delta\Phi_e(t) = \Phi_e^{drop}(t) + \frac{2RT(t)}{F}(1-t_+)\log_e{\left(\frac{c_{e,p}^{[1]}}{c_{e,n}^{[P]}}	\right)}
\end{align}
where, assuming that the ionic current $i_e(x,t)$ has a trapezoidal shape over the spatial domain \cite{Moura2017}, the electrolyte voltage drop $\Phi_e^{drop}(t)$ can be  approximated by
\begin{align}
\Phi_e^{drop}(t)\simeq  -\frac{I_{app}(t)}{2 P} \left( \phi_p(t) + 2\phi_s(t) + \phi_n(t) \right)
\end{align}
in which
\begin{subequations}
\label{eq:phi}
\begin{align}
\phi_p(t)&=\Delta x_p \sum_{k=1}^P \frac{2k-1}{\kappa(c_{e,p}^{[k]}(t),T(t))\epsilon_p^{p_p}} \\
\phi_s(t)&=\Delta x_s \sum_{k=1}^{P} \frac{1}{\kappa(c_{e,s}^{[k]}(t),T(t))\epsilon_s^{p_s}}\\
\phi_n(t)&=\Delta x_n \sum_{k=1}^{P} \frac{2P-2k+1}{\kappa(c_{e,n}^{[k]}(t),T(t))\epsilon_n^{p_n}}
\end{align}
\end{subequations}
where $\kappa(c_{e,j}^{[k]}(t),T(t))$ is the temperature-dependent electrolyte conductivity for the $k$th volume of the $j$th section, which is usually expressed as an empirically derived nonlinear function of the electrolyte concentration in that volume:
\begin{align}\label{eq:conductivity_fun}
\begin{split}
k(\gamma_j^{[k]}(t),T(t))=&\Big(0.2667 \left(\gamma_j^{[k]}(t)\right)^3 -1.2983  \left(\gamma_j^{[k]}(t)\right)^2\\& +1.7919 \gamma_j^{[k]}(t) + 0.1726\Big)e^{\frac{-E_{a,\kappa}}{RT(t)}}
\end{split}
\end{align}
where $\gamma_j^{[k]}(t)=10^{-3}c_{e,j}^{[k]}(t)$. The expression in \eqref{eq:conductivity_fun} is still referred to the case of  Kokam SLPB 75106100. 

% Note that, the expressions in \eqref{eq:phi} are computed considering different electrolyte conductivity values for each one of the discrete volumes constituting the cell sections. This approximantion provides better accuracy with respect to e.g. \cite{Moura2017}, in which the electrolyte conductivity is assumed constant over the spatial domain.

As mentioned above, this paper considers a lumped thermal model \cite{Perez2016, Perez2017} in which the temperature dynamics are given by
\begin{align}
C_{th} \dot T(t)= Q(t) - \frac{T(t)-T_{sink}}{R_{th}}
\end{align}
where $C_{th}$ is the  thermal capacity of the  cell and $R_{th}$ is the thermal resistance between the cell and the coolant, whose temperature is here assumed to be constant and equal to $T_{sink}$.  The heat generation term $Q(t)$ is due to the cell polarization and is described by
\begin{align}
Q(t)=\vert I_{app}(t) \vert \cdot \vert V(t) - (\bar U_{p}(t)-\bar U_{n}(t)) \vert . 
\end{align}
In order to simplify the notation in the next sections, define the variables for the average stoichiometry and concentration flux
\begin{subequations}\label{eq:simplified_notation}
\begin{align}
\overline{\theta}(t)&=\overline{\theta}_p(t)\\
\overline{q}(t)&=[\overline{q}_p(t),\,\overline{q}_n(t)]^\top
\end{align}
and express the vector of the electrolyte concentrations in the different finite volumes as
\begin{align}
c^e(t)=&[c_{e,p}^{[1]}(t),\,\cdots,\, c_{e,p}^{[P]}(t),\,c_{e,s}^{[1]}(t),\,\cdots,\, c_{e,s}^{[P]}(t),\,c_{e,n}^{[1]}(t),\,\cdots,\, c_{e,n}^{[P]}(t)]^\top.
\end{align}
\end{subequations}

\subsection{Model of a Battery Pack}\label{sub:battery_pack_model}
\begin{figure}[!htb]
\begin{center}
\includegraphics[width=\figsize\columnwidth]{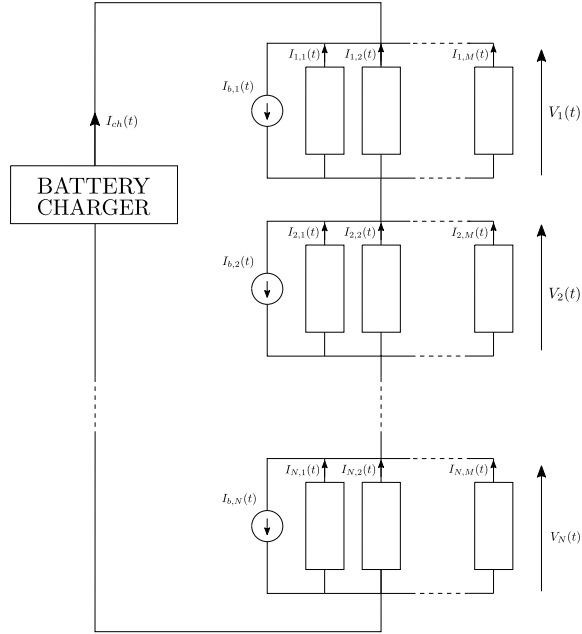}
\caption{Simplified circuital scheme.}
\label{fig:circuit_scheme}
\end{center}
\end{figure}

This subsection considers the battery pack configuration in Figure \ref{fig:circuit_scheme}, which consists of $N$ series-connected modules, each of which is constituted of $M$ parallel-connected cells. The total number of cells is given by $N_{cells}=N M$, with all cells modelled according to the equations in Section \ref{sub:model_of_a_single_cell}. The modelling of a lithium-ion battery pack as a set of connected electrochemical models constitutes a key point of this work with respect to the existing literature, in which the battery is usually described with a very simple equivalent circuit model. The use of a more accurate model is motivated by the fact that lithium-ion battery packs are complex systems which require suitable control strategies in order to guarantee the satisfaction of safety constraints on each single cell. However, the usage of an accurate model in optimal control has the disadvantage of high computational burden. This latter can be reduced by properly linearizing  the dynamics (see Section \ref{sec:Method}).

In the following, the indexes $i=1,\,\cdots,\,N$  and $j=1,\,\cdots,\,M$ which refer to the $j$th cell of the $i$th module. Moreover, the voltage, state of charge, temperature, and applied current of the cells are indicated respectively by $V_{i,j}(t)$, $SOC_{i,j}(t)$, $T_{i,j}(t)$, and $I_{i,j}(t)$, while the voltage of the $i$th module is defined as $V_i(t)$. 
The scheme in Figure \ref{fig:circuit_scheme} consists also of a supply circuitry which allows   charging of the battery (a battery charger) and balancing the energy stored in the different cells  (current generators). In particular, the battery charger supplies the current $I_{ch}(t)\geq0$, which can be assumed constant ($I_{ch}(t)=I_{ch}$), while the $N$ generators $I_{b,i}(t)\geq0,\, i=1,\,2,\,\cdots,\,N$ allow draining of the current which flows through the different modules. In the case of constant $I_{ch}$, these latter represent the system inputs and are fundamental in order to achieve an optimal current control of the different cells. Such current generators can be realized in practice in several ways. For instance, simple proportional-integral-derivative (PID) controllers can be used to regulate the value of a variable resistor in parallel to each module. As an alternative, a Pulse Width Modulation (PWM) approach or a method based on the cell bypass through active elements \cite{Pozzi2019} can be adopted. In this work, the heat generation due to the bypassing process is assumed to be negligible (see \cite{Gallardo-Lozano2014} for more details on the different supply and balancing circuitry schemes).
The battery pack is modelled as a system of DAEs with $N_{cells}$ algebraic variables ($I_{i,j}(t),\,i=1,\,\cdots,\,N,\,j=1,\,\cdots,\,M$) and corresponding $N_{cells}$ algebraic equations. Each module $i$, with $i=1,\,\cdots,N$, is described by $M$ algebraic equations:
\begin{subequations}
\begin{align}
%\begin{cases}
&V_{i,1}(t)=V_{i,2}(t)\\
&V_{i,2}(t)=V_{i,3}(t)\\
& \quad \quad \quad \vdots \\
&V_{i,M-1}(t)=V_{i,M}(t)\\
&I_{ch}(t)=-\sum_{j=1}^M I_{i,j}(t)+I_{b,i} \label{eq:kcl}
%\end{cases}
\end{align}
\end{subequations}
These equations correspond to the Kirchhoff's current law at each module and are written according to the conventions that the supply charger provides a positive current in order to charge the battery pack, the battery cells are charged by negative currents, and the bypassing system reduces the battery charging by draining positive currents.
The current generated by the battery charger is assumed to be completely bypassed through the generators $I_{b,i}$ when the $i$th module completes its charging procedure (i.e. $I_{b,i}(t)=I_{ch},\,t\geq \bar{t}_i$, where $\bar{t}_i$ is the time at which the $i$th module is completely charged). The heat exchange between the cells is also assumed to be negligible.  

% non parlo di top balancing perchè inutile
% non parlo neanche di autobilanciamento di celle in parallelo

\section{Sensitivity-Based Model Predictive Control}
\label{sec:Method}
This section proposes a sensitivity-based linear MPC (sMPC) for the control of general nonlinear continuous-time systems described by semi-explicit DAEs as in Section \ref{sub:nonlinear_dae_system}. The main advantage of this approach is the significant computational time reduction with respect to  nonlinear MPC (nMPC, see Section \ref{sub:nMPC}), while having comparable performance.

The sMPC strategy, which is presented in Section \ref{sub:linearization_dae_system}, relies on the model linearization around a nominal input trajectory, and is based on the computation of the sensitivities of states, outputs, and algebraic variables with respect to input perturbations. Such sensitivities are obtained by integrating additional continuous-time differential equations together with the model in \eqref{eq:NL_system}. Using this approach, the resulting linearized model provides higher accuracy than obtained with standard LTV approaches, in which the sensitivities are evaluated only at discrete time steps.
 
Next, Section \ref{sub:sMPC} describes how to use the sensitivity-based linearized model in an MPC framework. 
In order to address model mismatches in a practical implementation, the softening of the output constraints in sMPC and nMPC is described in Section \ref{sub:soft_constraints}.

\subsection{Nonlinear DAE system}\label{sub:nonlinear_dae_system}
Consider the continuous-time system of semi-explicit nonlinear DAEs described by
\begin{subequations}\label{eq:NL_system}
\begin{align}
\dot{x}(t)&=f(x(t),u(t),z(t))\\
0&=h(x(t),u(t),z(t)) \label{eq:algebraic_equation}\\
y(t)&=g(x(t),u(t),z(t)) \label{eq:output_map}
\end{align}
\end{subequations}
where $t \in \mathbb{R}$ is the time, $x(t) \in \mathbb{R}^n$ is the states vector, $u(t)\in \mathbb{R}^m$ is the control input, $z(t) \in \mathbb{R}^s$ is the vector of the algebraic variables, $y(t) \in \mathbb{R}^p$ is the output, $f:\mathbb{R}^{n\times m \times s}\rightarrow \mathbb{R}^n$ and $g:\mathbb{R}^{n\times m \times s}\rightarrow \mathbb{R}^p$ are the state and output functions respectively, and $h:\mathbb{R}^{n\times m \times s}\rightarrow \mathbb{R}^m$ specifies the set of algebraic constraints.
The system of DAEs in \eqref{eq:NL_system} is assumed to be index-1 \cite{Campbell1995}. Moreover, a digital controller is assumed to apply a piecewise constant input at the discrete times $t_k,\,k\in\mathbb{N}$ with sample time $T_s$. Within this context, define the generic input sequence applied in the time interval $[t_k,t_{k+H}]$, with $H\in\mathbb{N}$,  as 
\begin{align}
\mathbf{u}_{[t_k,t_{k+H}]}=\left[u^\top(t_k),\,u^\top(t_{k+1}),\,\cdots\,,u^\top(t_{k+H-1})\right]^\top
\end{align}
while the corresponding temporal evolution of states, algebraic variables, and outputs is obtained by integrating the equations in \eqref{eq:NL_system} over $[t_k,t_{k+H}]$, with initial condition $x(t_k)=x_k$, to give
\begin{subequations}\label{eq:sequences}
\begin{align}
\mathbf{x}_{[t_k,t_{k+H}]}&=\left[x^\top(t_k),\,x^\top(t_{k+1}),\,\cdots\,,x^\top(t_{k+H})\right]^\top\\
\mathbf{z}_{[t_k,t_{k+H}]}&=\left[z^\top(t_{k}),\,z^\top(t_{k+1}),\,\cdots\,,z^\top(t_{k+H})\right]^\top\\
\mathbf{y}_{[t_k,t_{k+H}]}&=\left[y^\top(t_{k}),\,y^\top(t_{k+1}),\,\cdots\,,y^\top(t_{k+H})\right]^\top
\end{align}
\end{subequations}

\subsection{Nonlinear MPC}\label{sub:nMPC} 
%mpc success in industry?? maybe better in the introduction
This section summarizes the main features of  nonlinear MPC, which is a control technique suitable for multivariable nonlinear systems in the presence of constraints. The formulation considers a continuous-time model as in \eqref{eq:NL_system} and a digital controller, which keeps the input constant over each sampling time.
Nonlinear MPC requires the solution of a finite-horizon optimal control problem at each time step  $t_{k_0}$, whose solution provides  the optimal control sequence $\mathbf{u^*}_{[t_{k_0},t_{k_0+H}]}$ over a  prediction horizon of $H$ steps, where 
\begin{align}
\mathbf{u^*}_{[t_{k_0},t_{k_0+H}]}=\left[{u^*}^\top(t_{k_0}),\,{u^*}^\top(t_{k_0+1}),\,\cdots\,,{u^*}^\top(t_{k_0+H-1})\right]^\top.
\end{align}
According to the \textit{receding horizon} paradigm, only the first element $u^*(t_{k_0})$ is applied to the system and the remaining future optimal moves  discarded. The optimization is then repeated at the next time step over a shifted prediction window, with the newly available measurements \cite{Bemporad1999}. 
 
The resulting optimization to be solved at each time $t_{k_0}$ is described below.
\begin{problem}\label{prob:problem}
Find the optimal input sequence $\mathbf{ u^*}_{[t_{k_0},t_{k_0+H}]}$ that solves 
\begin{align}
\mathbf{ u^*}_{[t_{k_0},t_{k_0+H}]}=\argmin_{\mathbf{u}_{[t_{k_0},t_{k_0+H}]}}J(t_{k_0})
\end{align}
for the cost function
\begin{align}\label{eq:total_cost}
J(t_{k_0})=\sum_{k=k_0}^{k_0+H}\Vert y(t_k)-y^{\text{ref}}\Vert^2_Q+\sum_{k=k_0}^{k_0+H-1}\Vert u(t_k)-u^{\text{ref}}\Vert^2_R +\textcolor{black}{J_{\text{reg}}(t_{k_0})}
\end{align}
in which $y^\text{ref} \in \mathbb{R}^p$ and $u^\text{ref} \in \mathbb{R}^m$ are the reference vectors for the output and for the input, respectively, and the matrices $Q \in \mathbb{R}^{p \times p}$ and $R \in \mathbb{R}^{m \times m}$ are design parameters which are weights in the MPC cost function, with $Q\geq0$ and $R > 0$. \textcolor{black}{The additional term $J_{\text{reg}}(t_{k_0})$ acts as a small regularization factor in order to avoid abrupt variations and spikes in the input control law and it is given by
\begin{align}\label{eq:regularization_term}
J_{\text{reg}}(t_{k_0})=\sum_{k=k_0}^{k_0+H-1}\Vert u(t_k)-u(t_{k-1})\Vert^2_{R_{\text{reg}}}
\end{align}
where $R_{\text{reg}}\in \mathbb{R}^{m \times m}$.}
Note that $y(t_k)\in \mathbb{R}^p$ is obtained by evaluating \eqref{eq:output_map} at discrete time instants $t_k$. 
The optimization is subject to the system dynamic in \eqref{eq:NL_system} and the constraints 
\begin{subequations}\label{eq:limits}
\begin{align}
u^{\text{lb}}\leq u(t_k)\leq u^{\text{ub}},&\quad k=k_0,\,k_0+1,\,\cdots,\,k_0+H-1\\
y^{\text{lb}}\leq y(t_k)\leq y^{\text{ub}},&\quad k=k_0,\,k_0+1,\,\cdots,\,k_0+H \label{eq:output_constraints}
%\\
%A\hat{y}(k)\leq b,&\quad k=k_0,\,\cdots,\,k_0+H 
%(\mathbf{E' NECESSARIA?})
\end{align}
\end{subequations}
where $u^{\text{lb}},\,u^{\text{ub}} \in \mathbb{R}^m$ and  $y^{\text{lb}},\,y^{\text{ub}} \in \mathbb{R}^p$  are the minimum and maximum allowable values of the inputs and outputs, respectively.
\end{problem}

\subsection{Sensitivity-Based Linearization Along a Nominal Trajectory}\label{sub:linearization_dae_system} 
The main drawback of the optimal control formulation in Section \ref{sub:nMPC} is its high computational cost, which comes from the use of a nonlinear model to predict the future system behaviour. This step involves the solution of a  nonlinear optimization at each time step whose complexity can be prohibitive for certain online control applications. 
As an alternative, we propose a sensitivity-based linearization of the system \eqref{eq:NL_system}, which can be exploited to provide a fast MPC solution. As in the previous sections,  the discussion is carried on by considering a digital controller which applies a  piecewise constant input at the discrete times  $t_k$.   
%the generic input sequence  defined over  a prediction horizon  of $H$ steps
%\begin{align}
%\mathbf{u}(k)=\left[u(k),\,u(k+1),\,\cdots\,,u(k+H-1)\right]
%\end{align}
%while the corresponding evolution of states, algebraic variables and outputs are given by
%\begin{subequations}\label{eq:sequences}
%\begin{align}
%\mathbf{x}(k)=\left[x(k),\,x(k+1),\,\cdots\,,x(k+H)\right]\\
%\mathbf{z}(k)=\left[z(k),\,z(k+1),\,\cdots\,,z(k+H)\right]\\
%\mathbf{y}(k)=\left[y(k),\,y(k+1),\,\cdots\,,y(k+H)\right]
%\end{align}
%\end{subequations}

Consider a nominal input signal $\overline{u}(t)$ and the nominal input sequence $\mathbf{\overline u}_{[t_k,t_{k+H}]}=[{\overline u}^\top(t_k),\,{\overline u}^\top(t_{k+1}),\,\cdots,\,{\overline u}^\top (t_{k+H-1})]^\top$ over the time window $[t_k,t_{k+H}]$. The corresponding  nominal trajectories for states, algebraic variables, and outputs are  $\mathbf{\overline x}_{[t_k,t_{k+H}]}$,  $\mathbf{\overline z}_{[t_k,t_{k+H}]}$, and $\mathbf{\overline y}_{[t_k,t_{k+H}]}$ (see \eqref{eq:sequences}). Define $S_x(t,t_{k})=\frac{\partial x(t)}{\partial u(t_k)}$, $S_z(t,t_{k})=\frac{\partial z(t)}{\partial u(t_k)}$, and $S_y(t,t_{k})=\frac{\partial y(t)}{\partial u(t_k)}$ which are the sensitivities of the states, algebraic variables, and outputs to a variation in the input $u(t_k)$ with respect to its nominal value $\overline u (t_k)$ at the discrete time $t_k$. In particular, the matrices $S_x(t,t_{k})\in \mathbb{R}^{n\times m}$, $S_z(t,t_{k})\in \mathbb{R}^{s\times m}$, and $S_y(t,t_{k})\in \mathbb{R}^{p\times m}$ are obtained by solving, together with \eqref{eq:NL_system}, the system of equations
\begin{subequations}\label{eq:sens_dynamics}
\begin{align}
\dot{S}_x(t,t_k)&=\overline{F}_x(t) S_x(t,t_k) +\overline{F}_z(t)S_z(t,t_k) + \overline{F}_u(t)\Delta_{t_k}(t) \\
0&=\overline{H}_x(t)S_x(t,t_k) + \overline{H}_z(t)S_z(t,t_k) + \overline{H}_u(t)\Delta_{t_k}(t) \label{eq:lin_algebraic_equation}	\\
S_y(t,t_k)&=\overline{G}_x(t)S_x(t,t_k) + \overline{G}_z(t)S_z(t,t_k) + \overline{G}_u(t)\Delta_{t_k}(t)
\end{align}
\end{subequations}
with initial condition $S_x(t_k,t_k)=\mathbf{0_{n \times m}}$ -- since we assume that \eqref{eq:NL_system} is a causal system -- and
\begin{subequations}\label{eq:jacobians}
\begin{align}
\overline{F}_\nu(t)=\nabla_{\nu} f(\overline{x}(t),\overline{z}(t),\overline{u}(t)),\quad \nu=\{x,\,z,\,u\}\\ 
\overline{H}_\nu(t)=\nabla_{\nu} h(\overline{x}(t),\overline{z}(t),\overline{u}(t)),\quad \nu=\{x,\,z,\,u\}\\
\overline{G}_\nu(t)=\nabla_{\nu} g(\overline{x}(t),\overline{z}(t),\overline{u}(t)),\quad \nu=\{x,\,z,\,u\}
\end{align}
\end{subequations}
where $\nabla_{\nu}$ is the Jacobian operator with respect to $\nu$ and $\Delta_{t_k}(t)=H(t-t_k)-H(t-t_k-T_s)$, with $H(t-t_k)$ being the unitary Heaviside step function
\begin{align}
H(t-t_k)=\begin{cases}
0,\quad t<t_k\\
1,\quad t\geq t_k
\end{cases}
\end{align}
%Note that in this work the initial conditions for the equations in \eqref{eq:sens_dynamics} are given by $S_x(t_k,t_k)=\mathbf{0_{n \times m}}$ but in general it may be necessary to obtain the initial conditions by directly applying the definition of sensitivity. 
%\begin{figure}[!htb]
%\begin{center}
%\includegraphics[width=0.8\columnwidth]{fig/Heaviside.eps}
%\caption{Representation of the signal $\Delta_{k_0}(t)$ as sum of unitary Heaviside functions.}
%\label{fig:Heaviside}
%\end{center}
%\end{figure}
Now consider a modified input sequence $\mathbf{\tilde{u}}_{[t_k,t_{k+H}]}=[\tilde u(t_k),\,\tilde u(t_{k+1}),\,\cdots,\,\tilde u (t_{k+H-1})]$  which can be obtained from the nominal input sequence by
\begin{align}
\mathbf{\tilde{u}}_{[t_k,t_{k+H}]}=\mathbf{\overline{u}}_{[t_k,t_{k+H}]}+\mathbf{\delta u}_{[t_k,t_{k+H}]}
\end{align}
where $\mathbf{\delta u}_{[t_k,t_{k+H}]}=[{\delta u}^\top(t_k),\,{\delta u}^\top(t_{k+1}),\,\cdots,\,{\delta u}^\top (t_{k+H-1})]^\top$ is the sequence of the input variations over the time window $[t_k,t_{k+H}]$. 
The sensitivity-based approximation of the states, algebraic variables, and output trajectories which correspond to the sequence $\mathbf{\tilde{u}}_{[t_k,t_{k+H}]}$ is given by \cite{Li1989,DeOliveira1995,Santin2016} 
\begin{subequations}\label{eq:linearized_model}
\begin{align}
\mathbf{\hat{x}}_{[t_k,t_{k+H}]}=\mathbf{\overline{x}}_{[t_k,t_{k+H}]}+\Pi^x_{[t_k,t_{k+H}]} \mathbf{\delta u}_{[t_k,t_{k+H}]}\\
\mathbf{\hat{z}}_{[t_k,t_{k+H}]}=\mathbf{\overline{z}}_{[t_k,t_{k+H}]}+\Pi^z_{[t_k,t_{k+H}]} \mathbf{\delta u}_{[t_k,t_{k+H}]}\\
\mathbf{\hat{y}}_{[t_k,t_{k+H}]}=\mathbf{\overline{y}}_{[t_k,t_{k+H}]}+\Pi^y_{[t_k,t_{k+H}]} \mathbf{\delta u}_{[t_k,t_{k+H}]}
\end{align}
\end{subequations} 
where the matrices $\Pi^\nu_{[t_k,t_{k+H}]}$, with $\nu=\{x,\,z,\,y\}$, are defined by
\begin{align}
\Pi^\nu_{[t_k,t_{k+H}]}=
\begin{bmatrix}
S_\nu(t_k,t_k)		&0					&\cdots	&0\\
S_\nu(t_{k+1},t_k)	&S_\nu(t_{k+1},t_{k+1})	&\cdots &0\\
S_\nu(t_{k+2},t_k)	&S_\nu(t_{k+2},t_{k+1})	&\cdots &0\\
\vdots 			&\vdots 			&\ddots &\vdots\\
S_\nu(t_{k+(H-1)},t_k)	&S_\nu(t_{k+(H-1)},t_{k+1})	&\cdots &S_\nu(t_{k+(H-1)},t_{k+(H-1)})\\
\end{bmatrix}
\end{align}
in which $S_\nu$, for $\nu=\{x,\,z,\,y\}$, is obtained by integrating the continuous-time system in \eqref{eq:sens_dynamics} together with \eqref{eq:NL_system}.
This step constitutes the main difference with respect to a standard LTV approach, for which the model sensitivities are computed only at discrete time instants, resulting in a loss of accuracy.

\subsection{Sensitivity-Based Linear MPC}\label{sub:sMPC} 
This section presents a model predictive control approach for the system \eqref{eq:NL_system} based on the linearized model  \eqref{eq:linearized_model}. Using the latter, the optimization can be formulated as a Quadratic Program (QP) which significantly reduces the computational cost compared to nonlinear MPC, thus enabling the use of the proposed strategy in on-line control applications. The objective function to be minimized at each time step $t_{k_0}$ is
\begin{align}\label{eq:total_cost_lin}
J_{lin}(t_{k_0})=\sum_{k=k_0}^{k_0+H}\Vert \hat y(t_k)-y^{\text{ref}}\Vert^2_Q+\sum_{k=k_0}^{k_0+H-1}\Vert \delta u(t_k)+\bar u(t_k)-u^{\text{ref}}\Vert^2_R\textcolor{black}{+J_{\text{reg}}(t_{k_0})}
\end{align}
and the resulting optimization is formulated below.
\begin{problem}\label{prob:problem_lin}
Find the optimal  sequence of the input variations  $\mathbf{ \delta u^*}_{[t_{k_0},t_{k_0+H}]}$ that solves
\begin{align}
\mathbf{ \delta u^*}_{[t_{k_0},t_{k_0+H}]}=\argmin_{\mathbf{\delta u}_{[t_{k_0},t_{k_0+H}]}}J_{lin}(t_{k_0})
\end{align}
for the cost function \eqref{eq:total_cost_lin}. The optimization is subject to the system dynamics in \eqref{eq:linearized_model} and the constraints 
\begin{subequations}\label{eq:limits_lin}
\begin{align}
u^{\text{lb}}\leq \tilde u(t_k)\leq u^{\text{ub}},&\quad k=k_0,\,k_0+1,\,\cdots,\,k_0+H-1\label{eq:box_constraints_lin}\\
y^{\text{lb}}\leq \hat y(t_k)\leq y^{\text{ub}},&\quad k=k_0,\,k_0+1,\,\cdots,\,k_0+H \label{eq:output_constraints_lin}
%\\
%A\hat{y}(k)\leq b,&\quad k=k_0,\,\cdots,\,k_0+H 
%(\mathbf{E' NECESSARIA?})
\end{align}
\end{subequations}
\end{problem}  
The optimal control sequence is obtained as
\begin{align}
\mathbf{u^*}_{[t_{k_0},t_{k_0+H}]}=\mathbf{\overline{u}}_{[t_{k_0},t_{k_0+H}]}+\mathbf{\delta u^*}_{[t_{k_0},t_{k_0+H}]}
\end{align}

The performance of the sensitivity-based linearization and the corresponding MPC algorithm are significantly affected by the choice of the nominal input sequence $\mathbf{\overline{u}}_{[t_{k_0},t_{k_0+H}]}$. 
Here we adopt the following choice. Suppose that, at the beginning, the nominal input sequence is not known. Then use the most likely input sequence as the initial guess, since no further information is available. Then, after each iteration of the MPC algorithm, the nominal input sequence is updated as
\begin{align}
\mathbf{\overline{u}}_{[t_{k_0+1},t_{k_0+H+1}]}=[\mathbf{{u}^*}^\top_{[t_{k_0+1},t_{k_0+H}]},\, \mathbf{{u}^*}^\top_{[t_{k_0+H-1},t_{k_0+H}]}]^\top
\end{align}
where $\mathbf{{u}^*}_{[t_{k_0},t_{k_0+H}]}$ is the optimal solution of the sMPC at the time $t_{k_0}$.
In case of highly nonlinear systems, the initialization of the nominal input sequence has to be done carefully to achieve a sufficiently accurate linearized model in the first iteration.

\subsection{Soft Constraints for Practical Implementation}\label{sub:soft_constraints} 
The constraints on the outputs for both the nonlinear \eqref{eq:output_constraints} and linearized \eqref{eq:output_constraints_lin} systems can be softened in order to deal with, in a practical implementation, possible model mismatches.  In particular, using a set of slack variables $\xi\in \mathbb{R}^p$, with $\xi\geq0$, the constraint in \eqref{eq:output_constraints} can be relaxed as
\begin{subequations}
\begin{align}
y^{\text{lb}}\leq y(t_k) +\xi(t_k),&\quad k=k_0,\,k_0+1,\,\cdots,\,k_0+H \\
y(t_k)-\xi(t_k)\leq y^{\text{ub}},&\quad k=k_0,\,k_0+1,\,\cdots,\,k_0+H 
\end{align}
\end{subequations}
and
\begin{subequations}
\begin{align}
y^{\text{lb}}\leq \hat y(t_k) +\xi(t_k),&\quad k=k_0,\,k_0+1,\,\cdots,\,k_0+H \\
\hat y(t_k)-\xi(t_k)\leq y^{\text{ub}},&\quad k=k_0,\,k_0+1,\,\cdots,\,k_0+H 
\end{align}
\end{subequations}
can be used for the constraint in \eqref{eq:output_constraints_lin}.

The cost functions for sMPC and nMPC are reformulated accordingly as 
\begin{subequations}
\begin{align}
J(t_{k_0})=&\sum_{k=k_0}^{k_0+H}\Vert  y(t_k)-y^{\text{ref}}\Vert^2_Q+\sum_{k=k_0}^{k_0+H-1}\Vert  u(t_k)-u^{\text{ref}}\Vert^2_R+\textcolor{black}{J_{\text{reg}}(t_{k_0})}+J_{s}(t_{k_0})\\
J_{lin}(t_{k_0})=&\sum_{k=k_0}^{k_0+H}\Vert \hat y(t_k)-y^{\text{ref}}\Vert^2_Q+\sum_{k=k_0}^{k_0+H-1}\Vert \delta u(t_k)+\bar u(t_k)-u^{\text{ref}}\Vert^2_R+\textcolor{black}{J_{\text{reg}}(t_{k_0})}\nonumber\\&+J_{s}(t_{k_0})
\end{align}
\end{subequations}
with
\begin{align}
J_{s}(t_{k_0})=\sum_{k=k_0}^{k_0+H-1}c^\top\xi(t_k)
\end{align}
where $c\in\mathbb{R}^p$ is a suitable vector of weights.
%Then, the time varying linearization of system \eqref{eq:NL_system} along the nominal trajectories  $\overline{x}(t)$   and  $\overline{z}(t)$, near the control  $\overline{u}(t)$ is given by
%\begin{subequations}\label{eq:lin_system}
%\begin{align}
%\dot{\delta x}(t)&=f_x(\overline{x}(t),\overline{u}(t),\overline{z}(t))\delta x + f_z(\overline{x}(t),\overline{u}(t),\overline{z}(t))\delta z +\nonumber\\&+ f_u(\overline{x}(t),\overline{u}(t),\overline{z}(t))\delta u  \\
%0&=h_x(\overline{x}(t),\overline{u}(t),\overline{z}(t))\delta x + h_z(\overline{x}(t),\overline{u}(t),\overline{z}(t))\delta z +\nonumber\\&+ h_u(\overline{x}(t),\overline{u}(t),\overline{z}(t))\delta u   \label{eq:lin_algebraic_equation}	\\
%\delta y(t)&=g_x(\overline{x}(t),\overline{u}(t),\overline{z}(t))\delta x + g_z(\overline{x}(t),\overline{u}(t),\overline{z}(t))\delta z +\nonumber\\&+ g_u(\overline{x}(t),\overline{u}(t),\overline{z}(t))\delta u   \label{eq:lin_output_map}
%\end{align}
%\end{subequations}

%---------------------------------------
\section{Results}
\label{sec:results}
This section evaluates the proposed methodology to the control of a lithium-ion battery pack. 
Section \ref{sub:adaptation_to_battery} introduces the optimal control problem, while the parameters and simulation settings are given in Section \ref{sub:parameters}. The simulation results are provided in Sections \ref{sub:compare_sens_nmpc}--\ref{sub:cccv}. Section \ref{sub:compare_sens_nmpc} compares the proposed sensitivity-based approach to nonlinear MPC, and a standard charging method, namely the Constant Current-Constant Voltage (CC-CV), is considered as a benchmark in Section \ref{sub:cccv}. Then, Section \ref{sub:scaling} analyzes how the  computational times of sMPC and nMPC grow with  increasing the number of cells. Finally, the methodologies are tested on a challenging scenario of the battery pack of an electric motorbike. 

\subsection{Optimal Control of a Lithium-Ion Battery}\label{sub:adaptation_to_battery}
In this section, the optimal control methods in Section \ref{sec:Method} are adapted to the optimal management of a lithium-ion battery, whose cells are arranged as in Figure \ref{fig:circuit_scheme}, with $N$ series modules of $M$ parallel-connected cells. The total number of cells is given by $N_{cells}=NM$. The objective of the control algorithm is to bring the state of charge of each cell as close as possible to  $100\%$ while satisfying all the safety constraints.
Each cell is modelled according to the equations described in Section \ref{sub:model_of_a_single_cell}, while the model of the whole battery pack is given in Section \ref{sub:battery_pack_model}. The variables are defined by
\begin{subequations}
\begin{align}
u(t_k)&=[I_{b,1}(t_k),\,I_{b,2}(t_k),\,\cdots,\,I_{b,N}(t_k)]^\top\\
x(t_k)&=[x^\top_{1,1}(t_k),\,x^\top_{1,2}(t_k),\,\cdots,\,x^\top_{N,M}(t_k)]^\top\\
z(t_k)&=[I_{1,1}(t_k),\,I_{1,2}(t_k),\,\cdots,\,I_{N,M}(t_k)]^\top\\
y(t_k)&=[y^\top_{1,1}(t_k),\,y^\top_{1,2}(t_k),\,\cdots,\,y^\top_{N,M}(t_k)]^\top
\end{align}
\end{subequations}
where $x_{i,j}(t_k)\in \mathbb{R}^{N_{x}}$ and $y_{i,j}(t_k)\in \mathbb{R}^{N_{y}}$ represent respectively the states and outputs of the $j$th cell of the $i$th module. The notation introduced in \eqref{eq:simplified_notation} implies that
\begin{subequations}
\begin{align}
x_{i,j}(k)&=\left[\bar \theta_{i,j}(k),\, \bar q^\top_{i,j}(k),\, {c^e}^\top_{i,j}(k),\, T_{i,j}(k) \right]^\top \\
y_{i,j}(k)&=\left[V_{i,j}(k),\,T_{i,j}(k),\,I_{i,j}(k),\, SOC_{i,j}(k)\right]^\top
\end{align}
\end{subequations}
where $\theta_{i,j}(k)$, $\bar q_{i,j}(k)$, and $c^e_{i,j}(k)$ refer to the stoichiometry, average concentration flux, and electrolyte concentration of the $j$th cell of the $i$th module. The elements of the vector $c^e_{i,j}(k)$ are the values of the electrolyte concentration along the spatial axis which is discretized according to the finite volume method. The resulting model is a semi-explicit continuous time system of DAEs (see \eqref{eq:NL_system}) where  $n=N_{cells}  N_{x}$, $m=N$, $s=N_{cells}$, and $p=N_{cells}  N_{y}$, with $N_{cells}=NM$ and in which $N_{x}=4+3P$ (for a $P$ number of finite volumes) and $N_{y}=4$ are the number of states and outputs of each single cell, respectively. 
The weighting matrices $Q \in \mathbb{R}_{\geq 0}^{p\times p}$ and $R\in \mathbb{R}_{> 0}^{m\times m}$ were specified to be diagonal
%\begin{align}
%Q=q \mathbb{I}_{p}
%\hspace{1cm}
%R=q \mathbb{I}_{m}
%\end{align}
\begin{align}
Q=\begin{bmatrix}
q	&0	&\cdots		&0\\
0	&q	&\cdots		&0\\
\vdots	&\vdots	&\ddots		&\vdots\\
0	&0	&\cdots		&q\\
\end{bmatrix},
\hspace{1cm}
R=\begin{bmatrix}
r	&0	&\cdots		&0\\
0	&r	&\cdots		&0\\
\vdots	&\vdots	&\ddots		&\vdots\\
0	&0	&\cdots		&r\\
\end{bmatrix}
\end{align}
where  $r \in \mathbb{R}_{> 0}$ and $q \in \mathbb{R}_{\geq 0}^{N_y \times N_y}$ is defined as 
\begin{align}
q=\begin{bmatrix}
q^{V}	&0		&0		&0\\
0		&q^{T}	&0  	&0\\
0		&0		&q^{I}	&0\\
0		&0		&0		&q^{SOC}
\end{bmatrix}.
\end{align}
The reference vector and the limits for the output vector, respectively, are given by 
\begin{subequations}
\begin{align}
y^{\text{ref}}&=[y^\top_{r},\,y^\top_{r},\,\cdots,\, y^\top_{r}]^\top\\
%u^{lb}&=[u^\top_{min},\,u^\top_{min},\,\cdots,\, u^\top_{min}]^\top\\
%u^{ub}&=[u^\top_{max},\,u^\top_{max},\,\cdots,\, u^\top_{max}]^\top\\
y^{lb}&=[y^\top_{min},\,y^\top_{min},\,\cdots,\, y^\top_{min}]^\top\\
y^{ub}&=[y^\top_{max},\,y^\top_{max},\,\cdots,\, y^\top_{max}]^\top
\end{align}
\end{subequations}
with  $y_r,\,y_{min},\,y_{max} \in \mathbb{R}^{N_y}$ defined by
\begin{subequations}
\begin{align}
%u_{min}&=MI_{min}\\
%u_{max}&=MI_{max}\\
y_{min}&=[V_{min},\, T_{min},\,I_{min},\, SOC_{min}]^\top\\
y_{max}&=[V_{max},\, T_{max},\,I_{max},\, SOC_{max}]^\top\\
y_{r}&=[V_{r},\, T_{r},\,I_{r},\, SOC_{r}]^\top
\end{align}
\end{subequations}
where $I_{min}$ and $I_{max}$ are the maximum and the minimum values of current that can flow through a single cell, $V_{min}$ and $V_{max}$ are the upper and lower bounds for the voltage, and $T_{min}$ and $T_{max}$, and $SOC_{min}$ and $SOC_{max}$, are the limits for the temperature and the state of charge respectively. Since the limits on the current flowing through each cell are explicitly considered in the optimization, lower and upper bounds on the input vector ($u^{lb}$ and $u^{ub}$) are not required. Finally, the reference values for voltage, temperature, current, and state of charge are denoted by $V_r$, $T_r$, $I_r$, and $SOC_r$ respectively. 

\begin{remark}
Although the internal states of each cell are not measurable in practice, all the relevant states are assumed available here. The use of observers goes beyond the scope of this work whose objective is to highlight the suitability of the sensitivity-based approach to real-time optimal control. For the design of observers for battery states, the interested reader can refer to e.g. \cite{Waag2014}. 
\end{remark}

\subsection{Model Parameters and Simulations Settings}\label{sub:parameters}
Here the virtual testbed considered in the simulations is described in detail in order to allow the presented results to be reproducible by others. The electrochemical parameters adopted for all the cells are those experimentally measured in \cite{Ecker2015,Ecker2015a} from a complete electrochemical characterization of a commercial cell (the Kokam SLPB 75106100). The thermal capacity, thermal resistance, and sink temperature are assumed equal to $C_{th}=4186\,$J/K, $R_{th}=169.5\,$K/W, and $T_{sink}=298.15\,$K, and the initial value for the temperature is set to $T^0=298.15\,$K for all cells. The initial electrolyte concentration and average concentration flux are assumed to start at equilibrium values $1000\,$mol/m$^3$ and zero respectively. The initial state of charge of the different cells, as well as the capacity and the SEI resistance, are extracted from a Gaussian distribution as
\begin{subequations}
\begin{align}
SOC^0_{i,j} &\in \mathcal{N}(SOC^0,\sigma^2_{SOC})\\
C^0_{i,j} &\in \mathcal{N}(C^0,\sigma^2_{C})\\
R_{sei,i,j}^0 &\in \mathcal{N}(R_{sei}^0,\sigma^2_{R_{sei}})
\end{align}
\end{subequations}
with $SOC^0=50\%$, 
$R_{sei}^0=15\,$m$\Omega$, 
and $C^0=7.5\,$Ah 
(i.e., $I_{1C}=7.5\,$A), 
while the standard deviations are 
$\sigma_{SOC}=10\%$, 
$\sigma_{R_{sei}}=0.75\,$m$\Omega$, and 
$\sigma_C=0.375\,$Ah.
%Moreover, the experimental data collected in \cite{Ecker2015} are fitted thus obtaining the following functions for the positive and negative open circuit potentials   
%\begin{subequations}
%\begin{align}\label{eq:OCPs_fun}
%\begin{split}
%\bar U_p(t)=&18.45\theta_p^6(t)-40.7 \theta_p^5(t)+20.94\theta_p^4(t)\\&+8.07\theta_p^3(t)-7.837\theta_p^2(t) + 0.02414\theta_p^1(t)+4.571
%\end{split}
%\\
%\bar U_n(t)=&    \frac{0.1261\theta_n(t)+0.00694}{\theta_n^2(t)+0.6995\theta_n(t)+0.00405}
%\end{align}
%\end{subequations}
%while the temperature-dependent  electrolyte conductivity function is expressed as follows
%\begin{align}\label{eq:conductivity_fun}
%\begin{split}
%k(\gamma_j^{[k]}(t),T(t))=&\Big(0.2667 \left(\gamma_j^{[k]}(t)\right)^3 -1.2983  \left(\gamma_j^{[k]}(t)\right)^2\\& +1.7919 \gamma_j^{[k]}(t) + 0.1726\Big)e^{\frac{-E_{a,\kappa}}{RT(t)}}
%\end{split}
%\end{align}
%where $\gamma_j^{[k]}(t)=10^{-3}c_{e,j}^{[k]}(t)$. 

The optimization settings are reported in Table \ref{tab:optimization_settings}, with the voltage, temperature, and current of each cell appearing in the constraint set but not weighted in the cost function ($q^V=0$, $q^T=0$, and $q^I=0$). The corresponding reference values $V_r$, $T_r$, and $I_r$ are set equal to zero. 

All the simulations were performed on a Windows 10 personal computer with 16 Gbytes of RAM and a 2.5 GHz i7vPro processor. The control problems were solved using \textit{CasADi} \cite{Andersson2012}, a symbolic framework for automatic differentiation. This software offers a Matlab interface for the Interior Point Optimization Method (IPOPT) \cite{Nocedal2006, Waechter2006} used for solving the optimizations, and for the SUNDIALS suite \cite{Hindmarsh2005} used for integrating the process dynamics. Moreover, \textit{CasADi} was used for the computation of the sensitivity matrices along the nominal trajectory. In order to provide a fair comparison between sMPC and nMPC, both of the underlying optimizations were solved using IPOPT.

% NMPC PARAMETERS ------------------------------------------------------------------------------------------------------------------
%\begin{table}[t!]
%\centering
%\renewcommand\arraystretch{1.3}
%\begin{tabular}{c|c|c|c}
%\toprule
%$\boldsymbol{P}$ 	& $\boldsymbol{H}$ 			&  $\boldsymbol{T_s}$ 				&  $\boldsymbol{SOC_r}$	  	  \\ \midrule
% $2$&     $3$			&     $40s$      	& $100\%$              	 	 \\ \toprule
%      $\boldsymbol{I_{min}}$  &   $\boldsymbol{V_{min}}$      				& $\boldsymbol{T_{min}}$                	& $\boldsymbol{SOC_{min}}$                         \\ \midrule
%        $-1.5I_{1C}$ &  $2.7 \ V$				& $253.15\ K$					& $0 \%$			 							
%        \\ \toprule
%      $\boldsymbol{I_{max}}$  &   $\boldsymbol{V_{max}}$      				& $\boldsymbol{T_{max}}$                	& $\boldsymbol{SOC_{max}}$                         \\ \midrule
%        $0\,A$ &  $4.2 \ V$				& $318.15\ K$					& $100 \%$			 	
%        \\ \toprule
%      $\boldsymbol{q^{V}}$  &   $\boldsymbol{q^{T}}$      				& $\boldsymbol{q^{SOC}}$                	& $\boldsymbol{r}$                         \\ \midrule
%        $0$ &  $0$				& $10^{-2}$					& $1.78\cdot 10^{-5}$								\\
%	  \bottomrule                 
%\end{tabular}
%\vspace{0.2cm}
%\caption{Parameters of the optimal control algorithms.}
%\label{tab:optimization_settings}
%\end{table}

\begin{table}[t!]
\centering
\renewcommand\arraystretch{1.3}
\begin{tabular}{c|c|c|c|c|c}
\toprule
$\boldsymbol{P}$ 	& $\boldsymbol{H}$ 			&  $\boldsymbol{T_s}$ 				&  $\boldsymbol{SOC_r}$	 &$\boldsymbol{I_{ch}}$  & $\boldsymbol{r}$	  \\ \midrule
 $2$&     $3$			&     $40\,$s      	& $100\%$              & $1.5M I_{1C}$ & $1.78 \times 10^{-5}$ 	 	 \\ \toprule
 $\boldsymbol{q^{V}}$  &   $\boldsymbol{q^{T}}$ &   $\boldsymbol{q^{I}}$      				& $\boldsymbol{q^{SOC}}$                	&
 $\boldsymbol{V_{min}}$    & $\boldsymbol{V_{max}}$   
 \\ \midrule
 $0$ &  $0$		&  $0$			& $10^{-2}$	 &$2.7\,$V		&$4.2\,$V		
         \\ \toprule

  $\boldsymbol{T_{min}}$               & $\boldsymbol{T_{max}}$                &     $\boldsymbol{I_{min}}$    &   $\boldsymbol{I_{max}}$  &
 $\boldsymbol{SOC_{min}}$                        &$\boldsymbol{SOC_{max}}$                 \\ \midrule        
    $253.15\,$K &$318.15\,$K  & $-1.5I_{1C}$ & $0\,$A &$0\%$ &$100\%$
     	\\
	  \bottomrule                 
\end{tabular}
\vspace{0.2cm}
\caption{Parameters of the optimal control algorithms.}
\label{tab:optimization_settings}
\end{table}

%% ----------------------
\subsection{Comparison Between sMPC and nMPC}\label{sub:compare_sens_nmpc}
\begin{figure}[!tb]
\begin{center}
\subfigure[State of charge]{
\includegraphics[width=0.47\columnwidth]{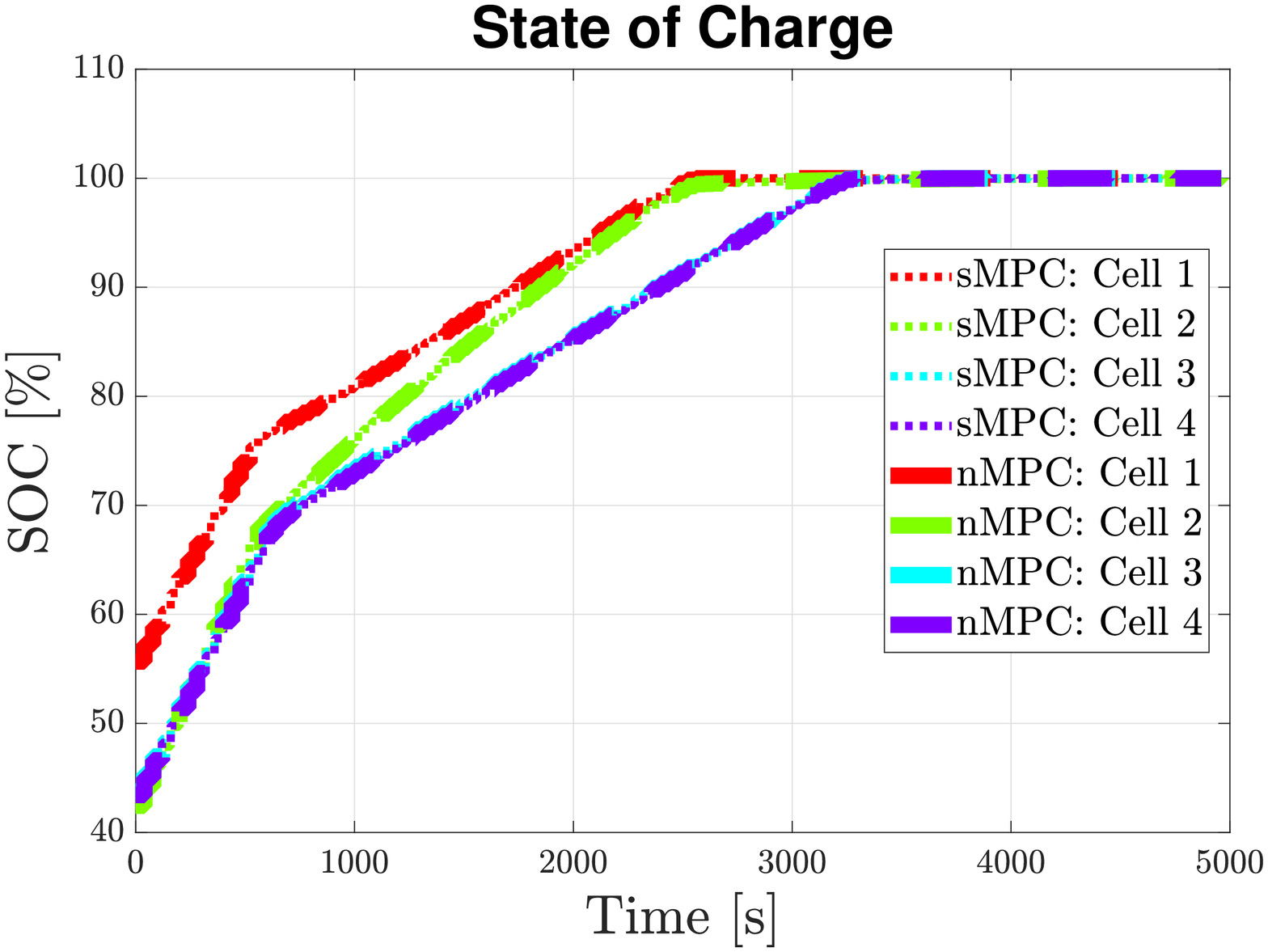}\label{fig:soc}}
\subfigure[Voltage]{
\includegraphics[width=0.47\columnwidth]{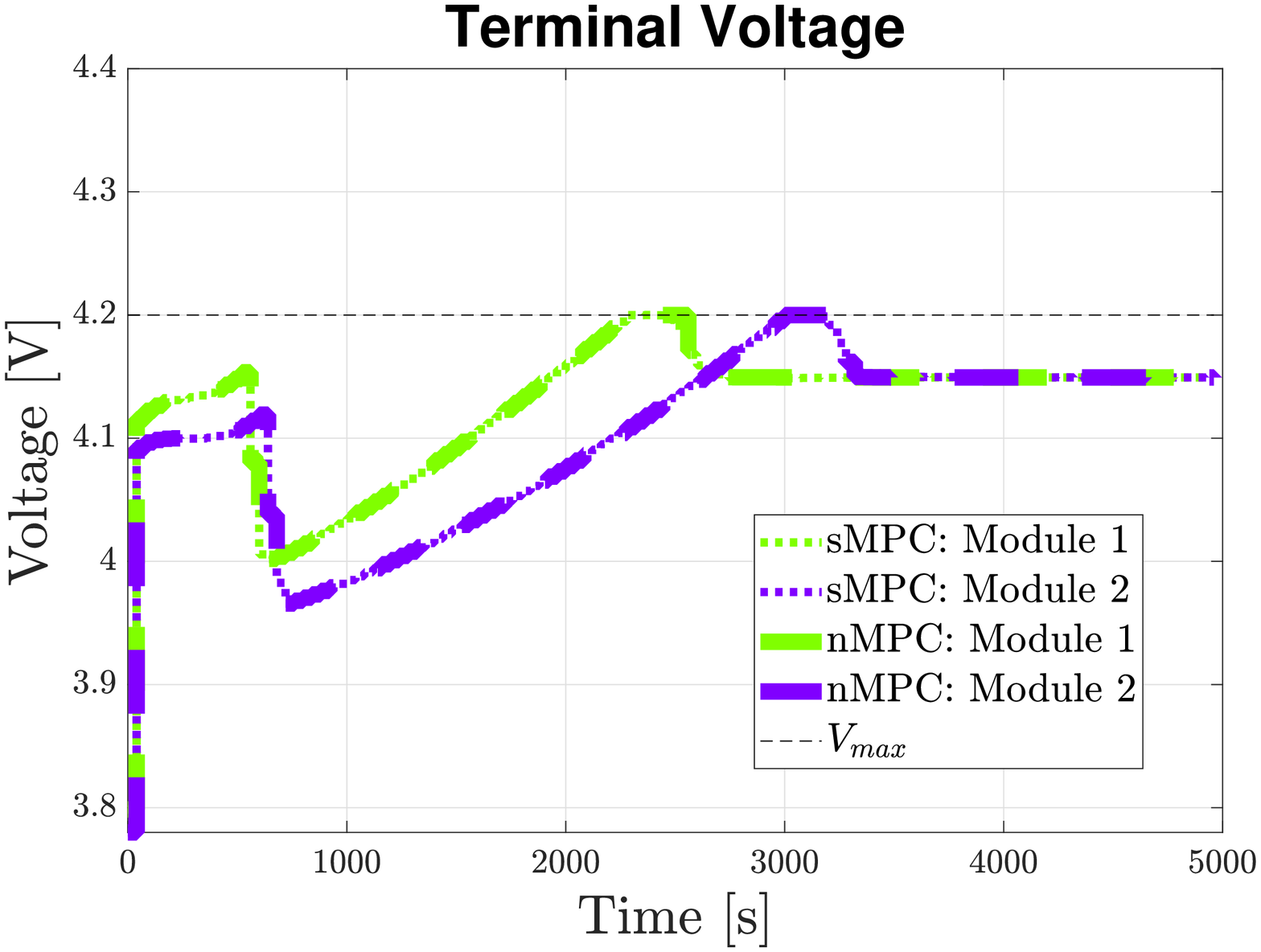}\label{fig:voltage}}
\caption{Temporal evolution of the state of charge and voltage for sMPC and nMPC. Only the voltage of the two modules is shown in (b) since all the cells of a particular module present the same voltage due to the parallel connection.}
\label{fig:soc_and_voltage}
\end{center}
\end{figure}

\begin{figure}[!tb]
\begin{center}
\subfigure[Temperature]{
\includegraphics[width=0.47\columnwidth]{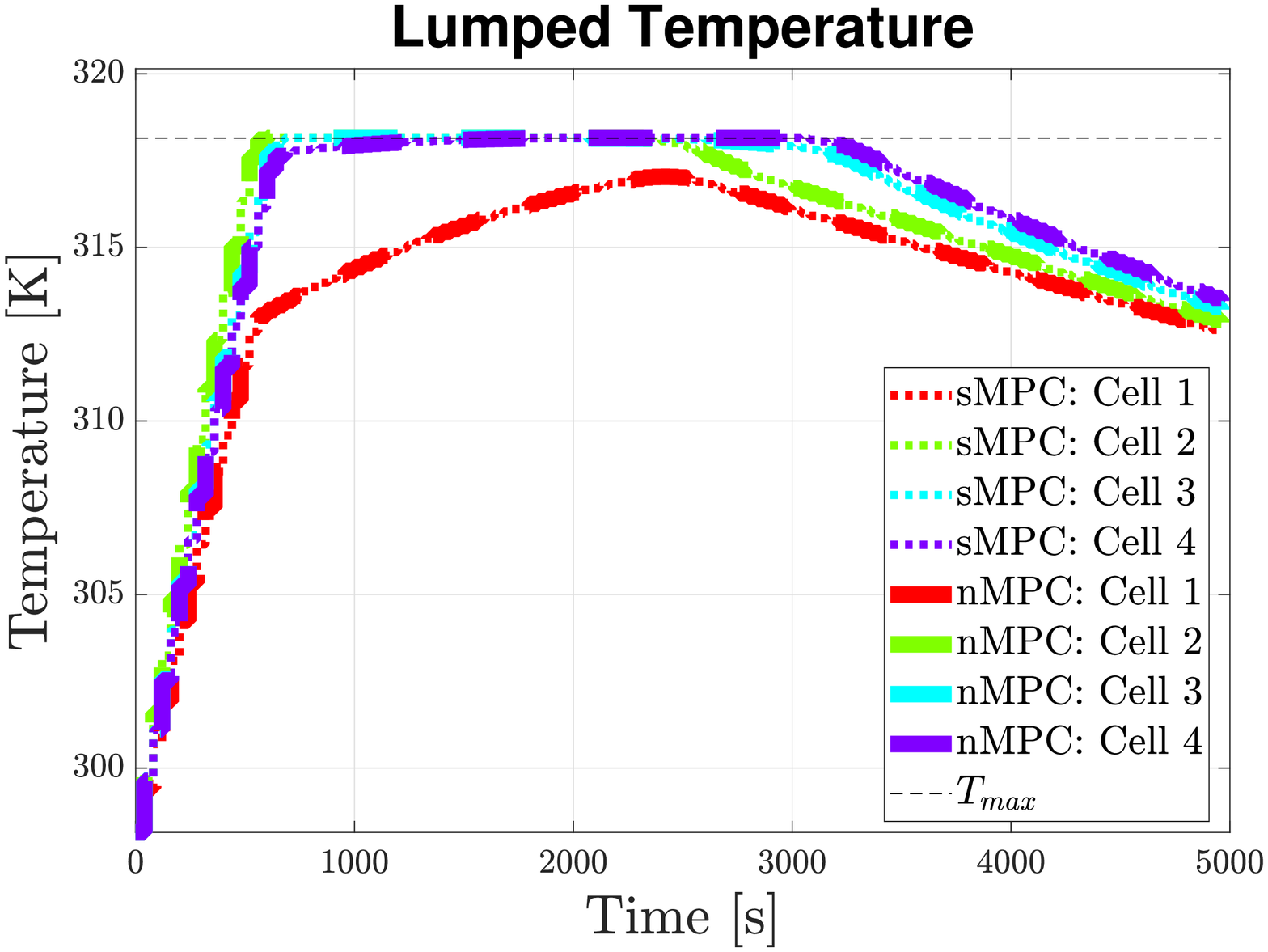}\label{fig:temperature}}
\subfigure[Current]{
\includegraphics[width=0.47\columnwidth]{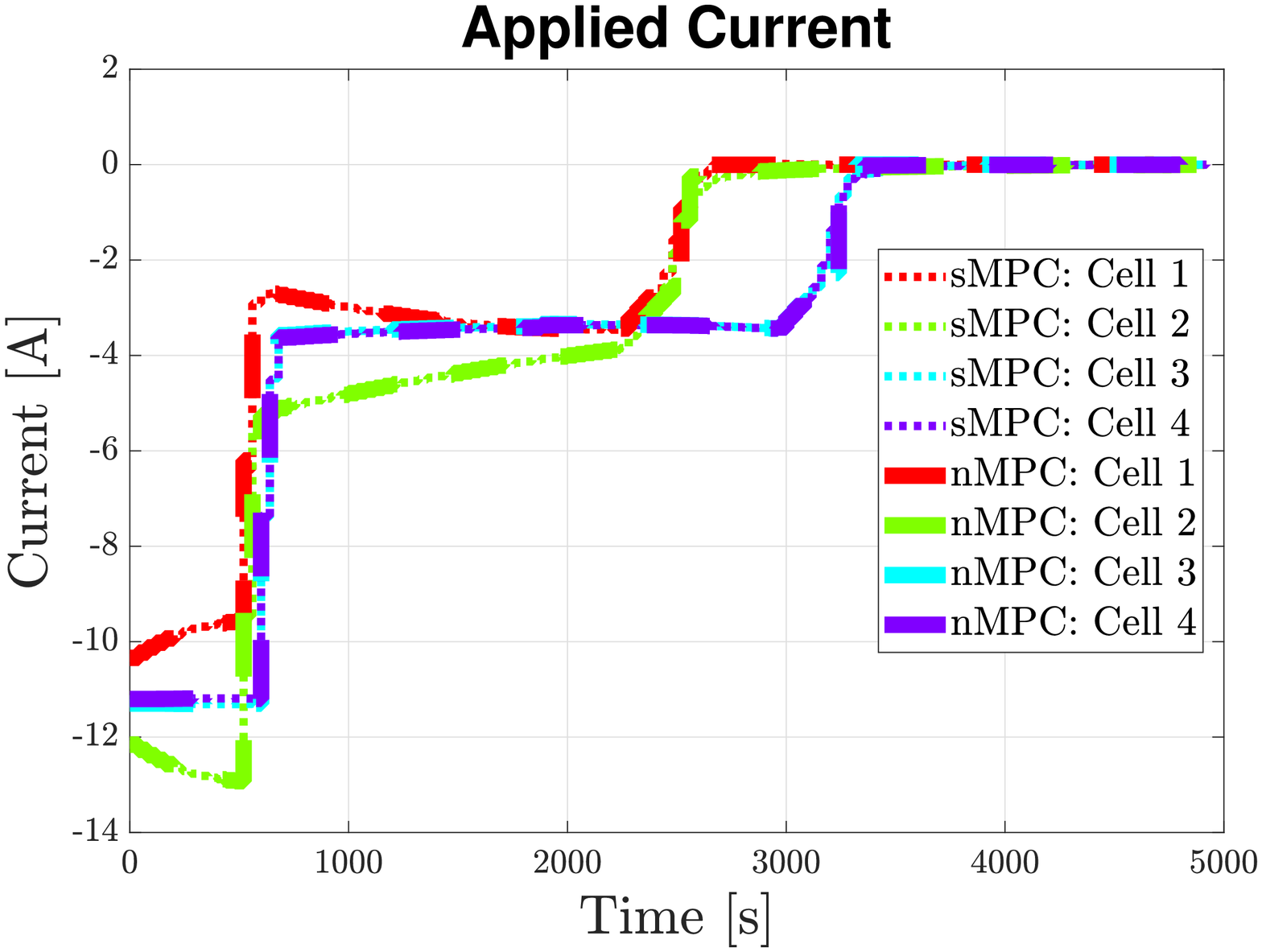}\label{fig:current}}
\caption{Temporal evolution of the temperature and current for sMPC and nMPC.}
\label{fig:temperature_and_current}
\end{center}
\end{figure}

\begin{figure}[!tb]
\begin{center}
\subfigure[Drained current]{
\includegraphics[width=0.47\columnwidth]{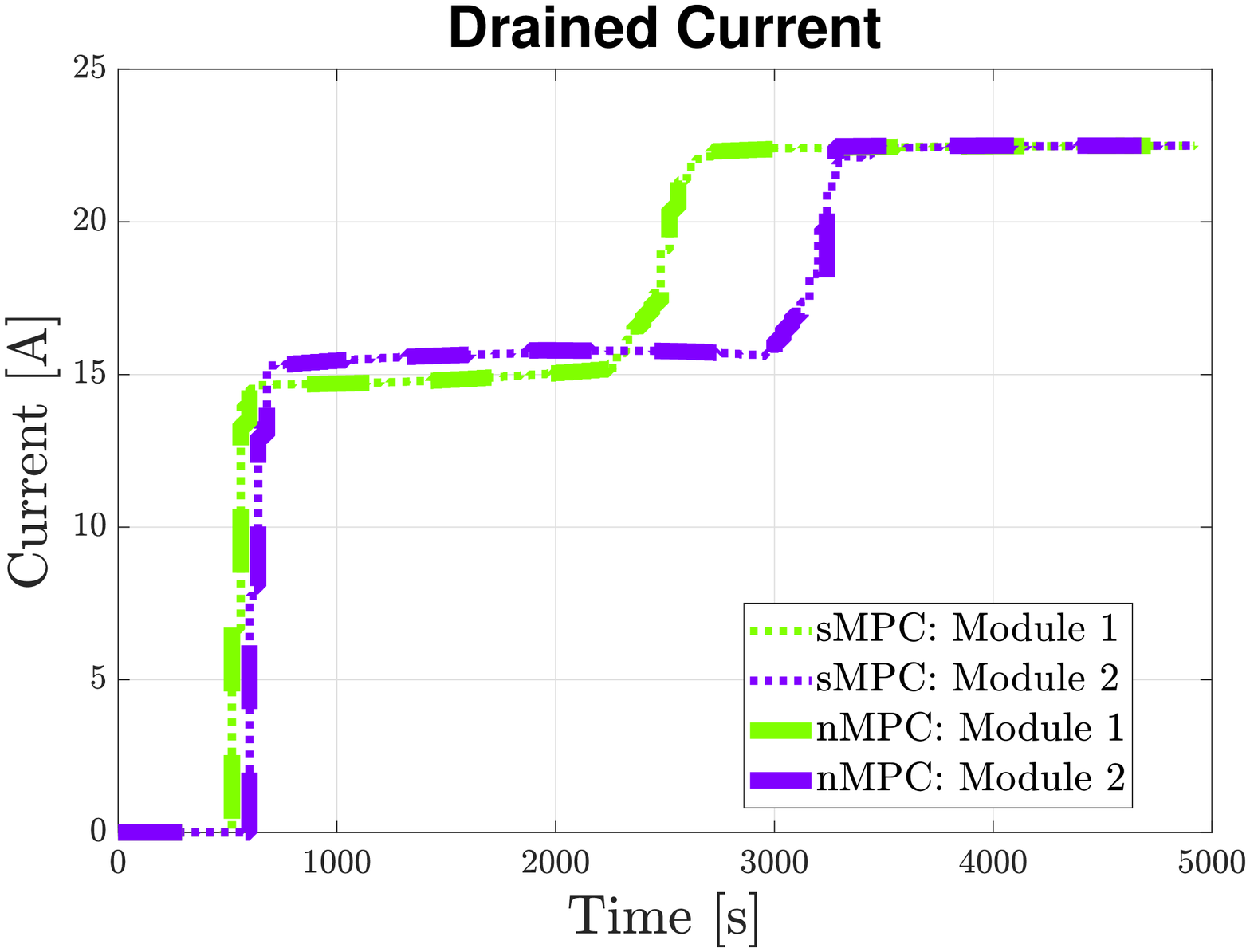}\label{fig:input}}
\subfigure[Computational time]{
\includegraphics[width=0.47\columnwidth]{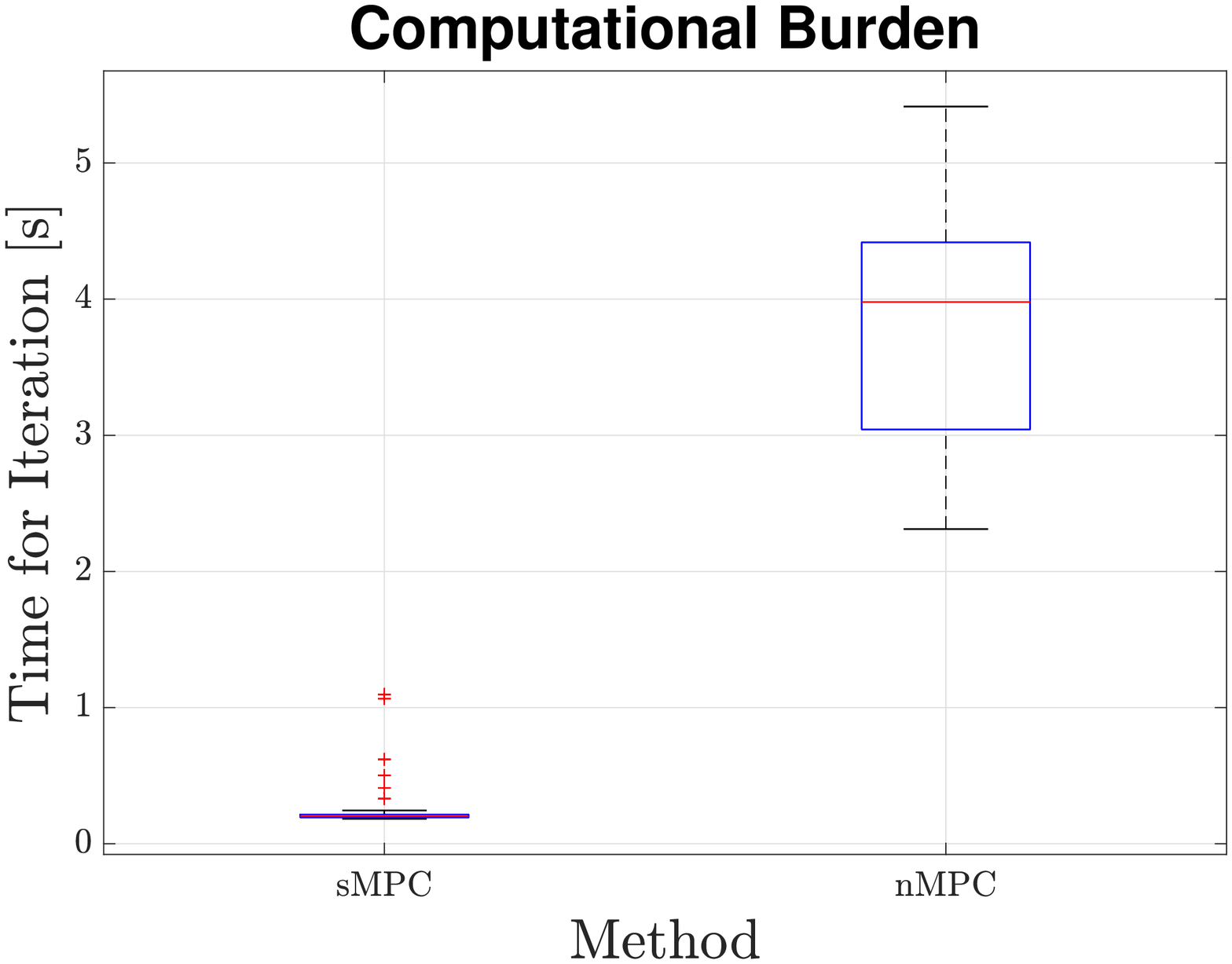}\label{fig:computational_time}}
\caption{Temporal evolution of the drained current  for the two modules and the computational times for sMPC and nMPC.} %RDB: the plot top title on the left is ``Bypassing Currents'' whereas the heading is ``Drained currents''. Remove the top titles for all figures.
\label{fig:input_and_time}
\end{center}
\end{figure}

This section considers a battery pack composed by $N=2$ modules with $M=2$ parallel connected cells for each module. The performance of the proposed sensitivity-based MPC is compared with the nonlinear MPC as the benchmark for the parameters of each cell reported in the previous section. The temporal evolution of the states and outputs obtained by sMPC (dotted line) and nMPC (dashed line) are very similar (see Figures 2a--4a), with nearly complete overlap. For both MPC formulations, the state of charge for all of the cells reach the desired target of $100\%$ within $3500\,$s (Figure \ref{fig:soc}). The constraints on the voltage and temperature for all of the battery cells are satisfied for all time (Figures \ref{fig:voltage} and \ref{fig:temperature}). The current flowing in the different cells are nearly identical for the two MPC formulations (Figure \ref{fig:current}),  and the input actions are nearly identical for each module (Figure \ref{fig:input}), consisting of the drained currents $I_{b,1}$ and $I_{b,2}$.

Since the main goal of this paper is to develop a control algorithm that achieves high performance while having low enough on-line computational cost being implementable in real time, the mean computational times needed by the two methods to compute the optimal control sequence at each time step are reported in Figure \ref{fig:computational_time}. While having very similar closed-loop performance, the on-line computational cost of the sensitivity-based MPC is significantly lower than for nMPC, motivating the use of sMPC in the context of real-time control of battery packs. Also, the on-line computational time of nMPC is highly variable while being nearly deterministic for sMPC, as its optimization is a quadratic program whose computational cost for solution is weakly dependent on the values of its parameters. Having low variability in its on-line computational cost is another desirable feature of sMPC.

%----------- SOC and VOLTAGE 

%\begin{figure}[!htb]
%\begin{center}
%\includegraphics[width=\figsize\columnwidth]{fig/mpc/soc.eps}
%\caption{State of charge evolution.}
%\label{fig:soc}
%\end{center}
%\end{figure}
%
%
%
%\begin{figure}[!htb]
%\begin{center}
%\includegraphics[width=\figsize\columnwidth]{fig/mpc/voltage.eps}
%\caption{Voltage evolution.}
%\label{fig:voltage}
%\end{center}
%\end{figure}

%----------- TEMPERATURE AND CURRENT
%
%\begin{figure}[!htb]
%\begin{center}
%\includegraphics[width=\figsize\columnwidth]{fig/mpc/temperature.eps}
%\caption{Temperature evolution.}
%\label{fig:temperature}
%\end{center}
%\end{figure}
%
%
%\begin{figure}[!htb]
%\begin{center}
%\includegraphics[width=\figsize\columnwidth]{fig/mpc/current.eps}
%\caption{Current evolution.}
%\label{fig:current}
%\end{center}
%\end{figure}

% -------------- INPUT and TIME
%
%
%\begin{figure}[!htb]
%\begin{center}
%\includegraphics[width=\figsize\columnwidth]{fig/mpc/input.eps}
%\caption{Inputs evolution.}
%\label{fig:input}
%\end{center}
%\end{figure}
%
%
%
%\begin{figure}[!htb]
%\begin{center}
%\includegraphics[width=\figsize\columnwidth]{fig/mpc/computational_time.eps}
%\caption{Computational time comparison.}
%\label{fig:computational_time}
%\end{center}
%\end{figure}

%-----------------------------------------

\subsection{Standard CC-CV Method}\label{sub:cccv}
To demonstrate the need for an optimal management of a lithium-ion battery pack that is able to take into account input and output constraints, a standard charging method -- namely the Constant Current-Constant Voltage (CC-CV) -- is applied to the same configuration considered in Section \ref{sub:compare_sens_nmpc}. The CC-CV method is composed of two phases. In the first phase, a constant charging current $I_{cc}$ is applied to the series connected modules. This phase ends separately for the different modules, as soon as their voltage reaches a predefined  threshold $V_{th}$ (that in this case is assumed to be equal to $4.15\,$V, which corresponds to the OCPs difference value at $100\%$ of SOC). Note that this value can be lower than the maximum allowed voltage specified by the cell data-sheet. The second phase consists of a separate constant voltage charging of the different modules with a voltage generator of $V_{cv}=V_{th}$. The charging procedure is completed when the maximum current flowing through the different modules achieves the threshold $I_{th}$ (which in this case is assumed to be equal to $0.1MI_{1C}$). In order to implement such a procedure, the battery pack needs to be equipped with a system of switches which allows to commute from the first phase to the second phase of the CC-CV.

\begin{figure}[!tb]
\begin{center}
\subfigure[State of charge]{
\includegraphics[width=0.47\columnwidth]{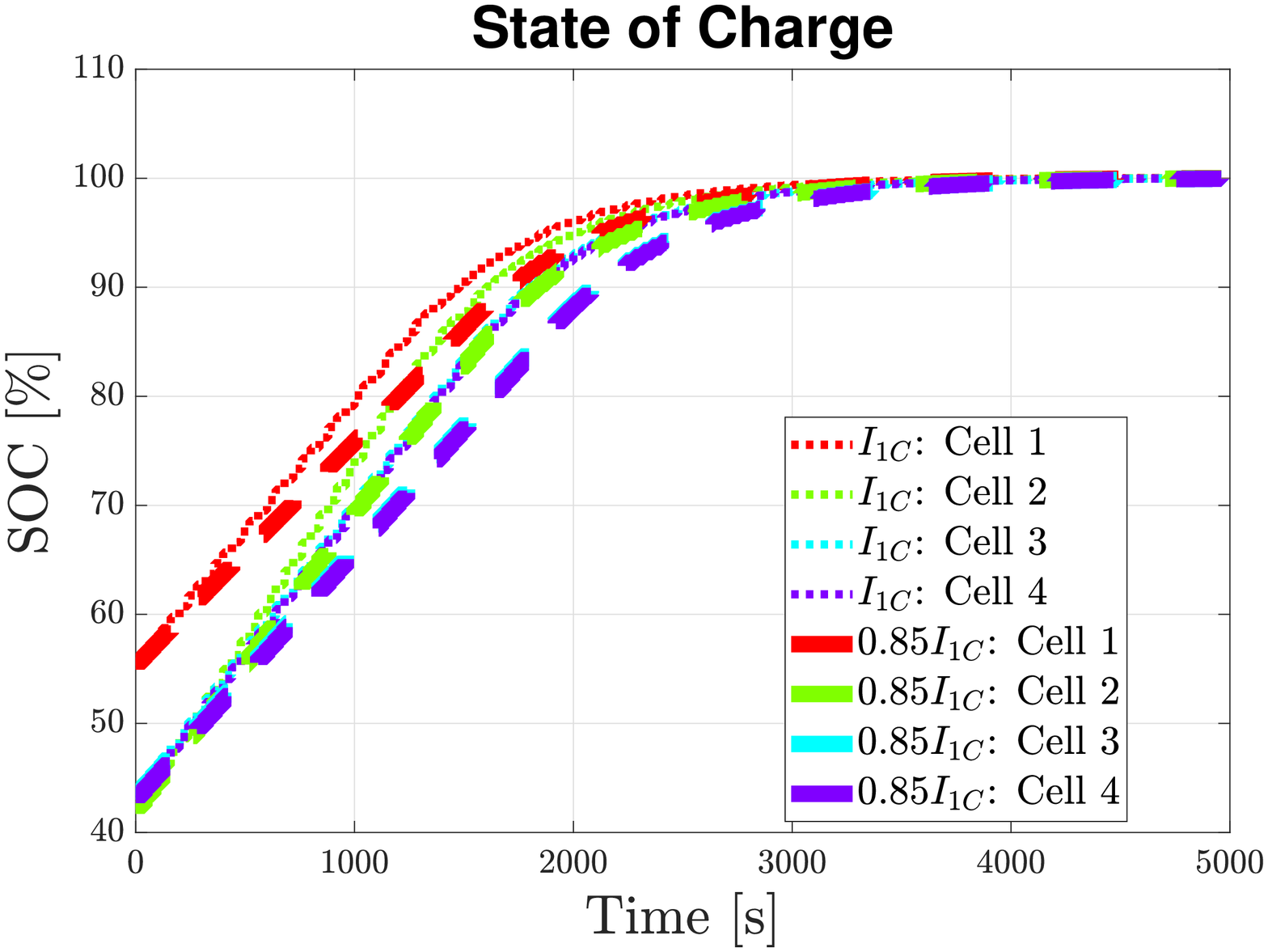}\label{fig:soc_cccv}}
\subfigure[Voltage]{
\includegraphics[width=0.47\columnwidth]{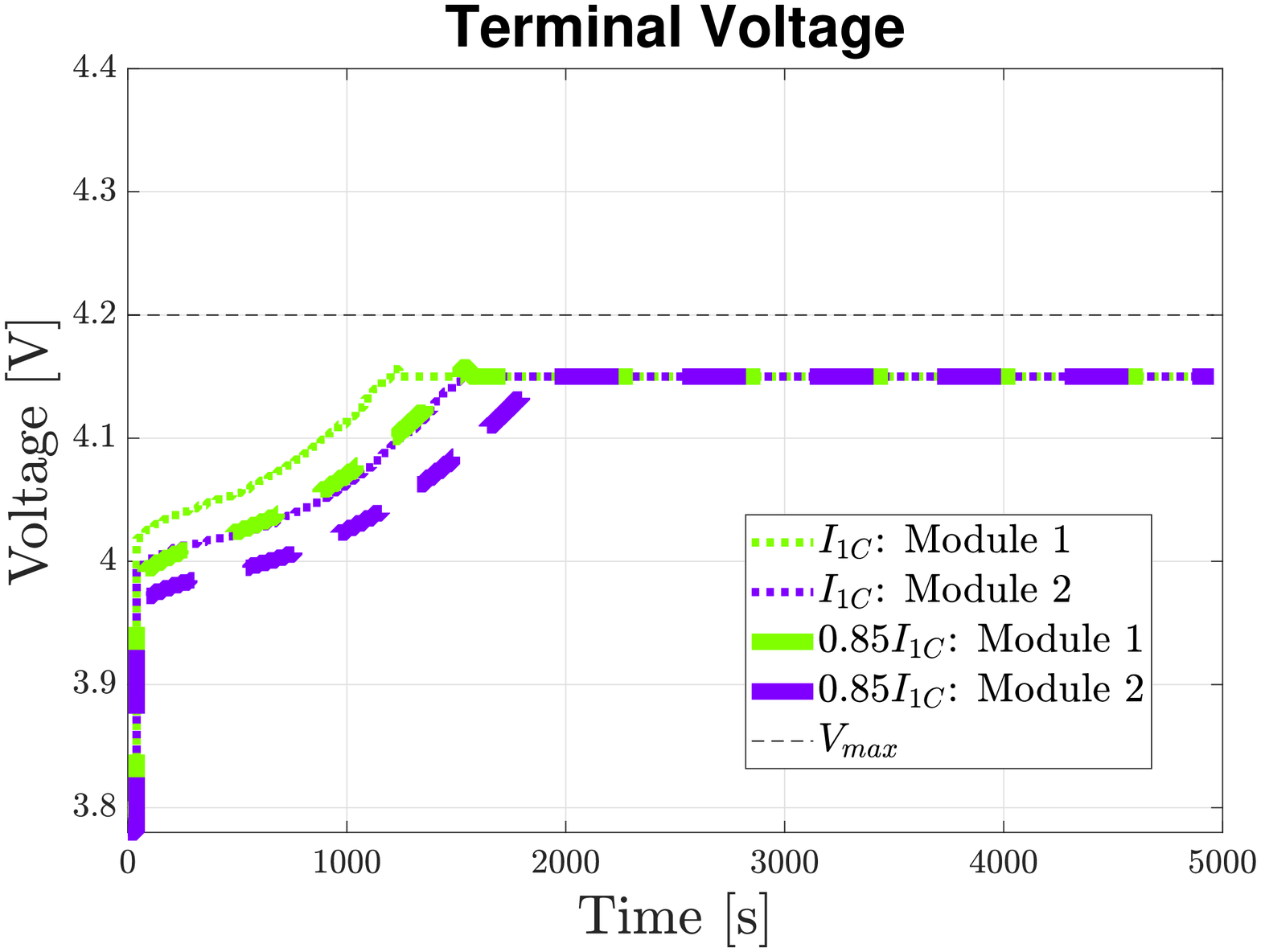}\label{fig:voltage_cccv}}
\caption{Temporal evolution of the state of charge and voltage for CC-CV charging for two values for the value of the constant current}
\label{fig:soc_and_voltage_cccv}
\end{center}
\end{figure}

\begin{figure}[!tb]
\begin{center}
\subfigure[Temperature]{
\includegraphics[width=0.47\columnwidth]{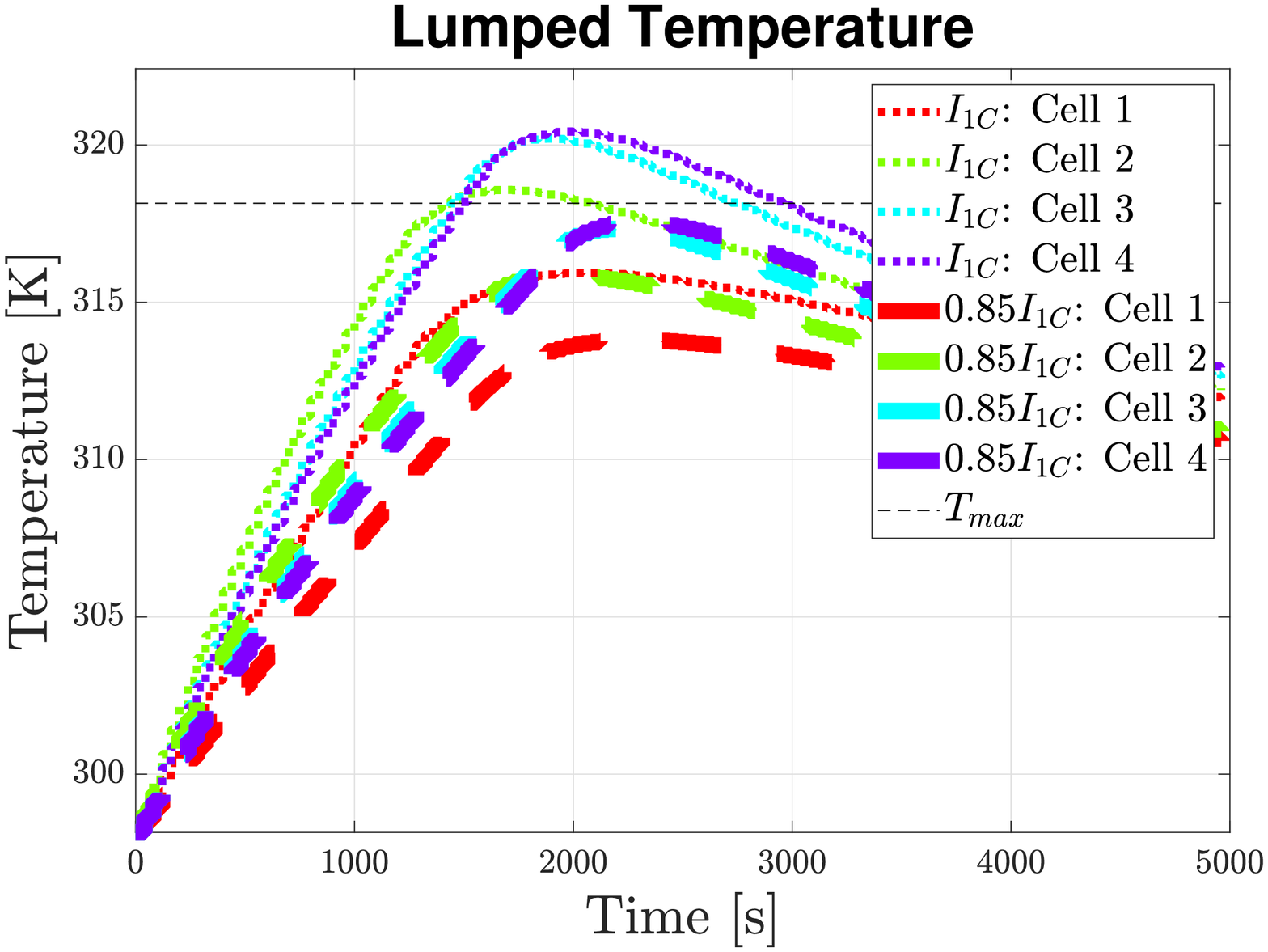}\label{fig:temperature_cccv}}
\subfigure[Current]{
\includegraphics[width=0.47\columnwidth]{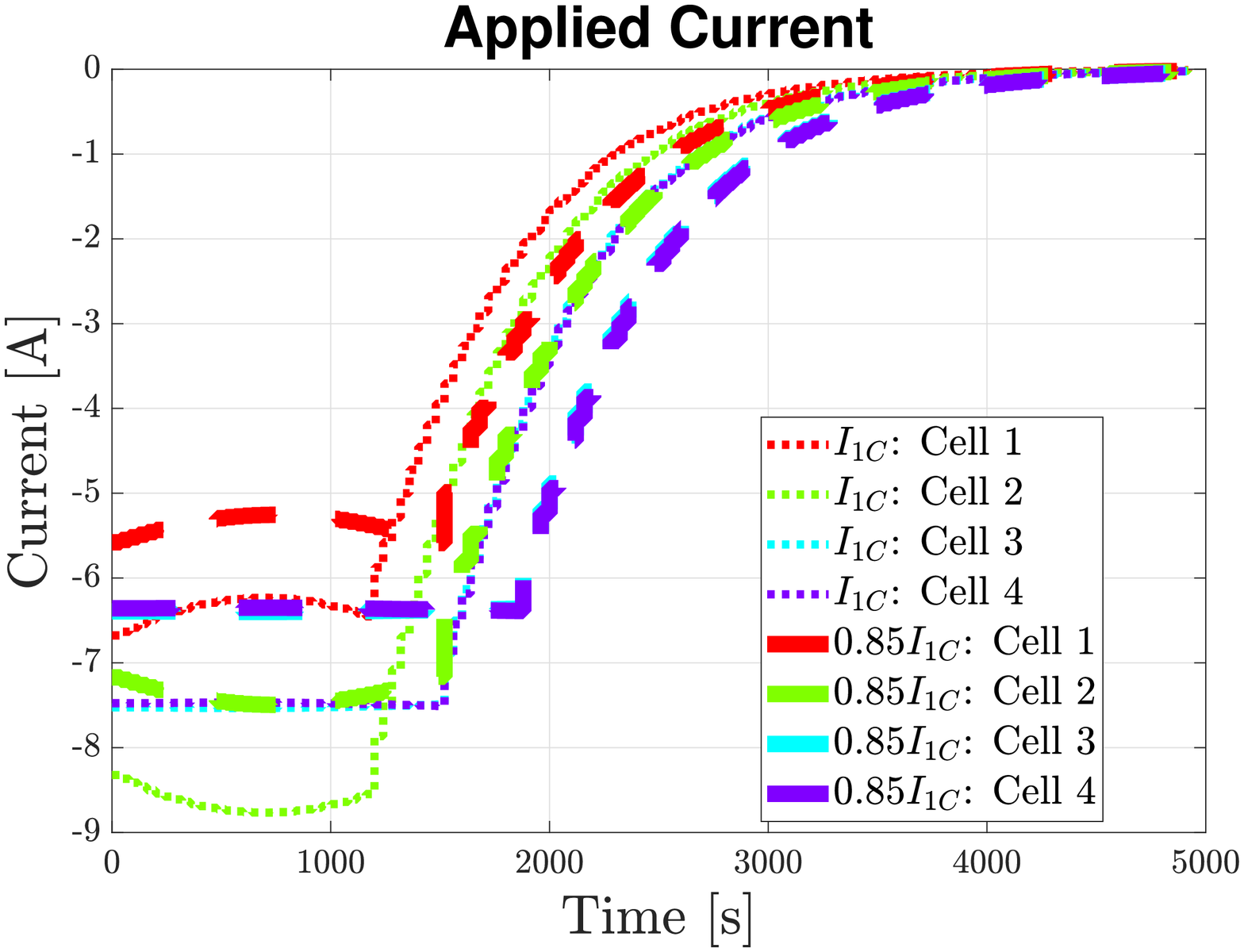}\label{fig:current_cccv}}
\caption{Temporal evolution of the temperature and applied current (the sum of the current for the cells of each module produces the CC-CV profile) for CC-CV charging for two values for the constant current.}
\label{fig:temperature_and_current_cccv}
\end{center}
\end{figure}

For the CC-CV charging protocols, the temporal evolution of the state of charge, voltage, temperature, and currents applied to the different cells are shown in Figures \ref{fig:soc_cccv}, \ref{fig:voltage_cccv}, \ref{fig:temperature_cccv}, and  \ref{fig:current_cccv}. Two scenarios were considered which differ in the value of the current applied during the constant current phase. The constant current of $I_{cc}=MI_{1C}$ (dotted line) results in high temperature constraint violation in most of the cells, with a charging time of $3960\,$s. On the other hand, for the lower constant current (dashed line) of $I_{cc}=0.85MI_{1C}$, the temperature constraint is satisfied for each cell, but the charging time increases significantly ($4360\,$s). Moreover, the value of the constant current $I_{cc}$ which guarantees the satisfaction of the temperature constraint must be found experimentally and can change according to the external environment conditions as well as with the battery ageing and degradation. %RDB: The text and Figures 5 and 6 indicate that the slower CC-CV is 0.85 I_1c whereas Table 2 indicates that the slower CC-CV is 0.65 I_1c

Table \ref{tab:evaluation_criteria} compares the charging time, computational time, and maximum temperature and voltage for the sMPC, nMPC, and two CC-CV charging protocols. The MPC algorithms have the same charging times and maximum temperature and voltage, while sMPC required only about 6\% of the on-line computational time.

%----- SOC and Voltage

%
%
%\begin{figure}[!htb]
%\begin{center}
%\includegraphics[width=\figsize\columnwidth]{fig/mpc/cc_cv_soc.eps}
%\caption{CC-CV: State of charge evolution.}
%\label{fig:soc_cccv}
%\end{center}
%\end{figure}
%
%
%
%\begin{figure}[!htb]
%\begin{center}
%\includegraphics[width=\figsize\columnwidth]{fig/mpc/cc_cv_voltage.eps}
%\caption{CC-CV: Voltage evolution.}
%\label{fig:voltage_cccv}
%\end{center}
%\end{figure}

%---------- TEMPERATURE and CURRENT
%
%\begin{figure}[!htb]
%\begin{center}
%\includegraphics[width=\figsize\columnwidth]{fig/mpc/cc_cv_temperature.eps}
%\caption{CC-CV: Temperature evolution.}
%\label{fig:temperature_cccv}
%\end{center}
%\end{figure}
%
%
%\begin{figure}[!htb]
%\begin{center}
%\includegraphics[width=\figsize\columnwidth]{fig/mpc/cc_cv_current.eps}
%\caption{CC-CV: Current evolution.}
%\label{fig:current_cccv}
%\end{center}
%\end{figure}

\begin{table}[t!]
\centering
\renewcommand\arraystretch{1.3}
\begin{tabular}{c|c|c|c|c}
\toprule
					& \multirow{2}{*}{sMPC}	& \multirow{2}{*}{nMPC}   		&  CC-CV & CC-CV 	  \\
& & 	&($I_{1C}$) & ($0.85I_{1C}$)	\\
					 \toprule
Charging Time 		&     $3280\,$s			& $3280\,$s  		& $3960\,$s    	   	& $4360\,$s         	 	 \\ \midrule
Comp. Time 	&     $0.24\,$s			& $3.91\,$s  		& --   	   			& --       	 	 \\ \midrule
Max. Temp. &     $318.15\,$K		&  $318.15\,$K  &  $320.42\,$K  	   	& $317.55\,$K    	 	 \\ \midrule
Max. Voltage 	&     $4.2\,$V			&  $4.2\,$V  	&  $4.15\,$V  	   	& $4.15\,$V   	 	 \\  \bottomrule                 
\end{tabular}
\vspace{0.2cm}
\caption{Comparison of charging time, computational time and maximum temperature and voltage reached for sMPC, nMPC, and CC-CV charging.}
\label{tab:evaluation_criteria}
\end{table}

%--------------------------------------------------------------------------
\subsection{Scaling of the Computational Time for Increasing Number of Cells}\label{sub:scaling}
The above simulations showed that the sMPC has much lower on-line computational cost than nMPC for a battery pack of 4 cells. Figure 7 displays the mean computational time for sMPC and nMPC for an increasing the number of series and parallel connections. sMPC is about an order of magnitude faster than nMPC, which is a significant savings in on-line computational cost when dealing with large battery packs. 

\begin{figure}[!tb]
\begin{center}
\subfigure[sMPC]{
\includegraphics[width=0.47\columnwidth]{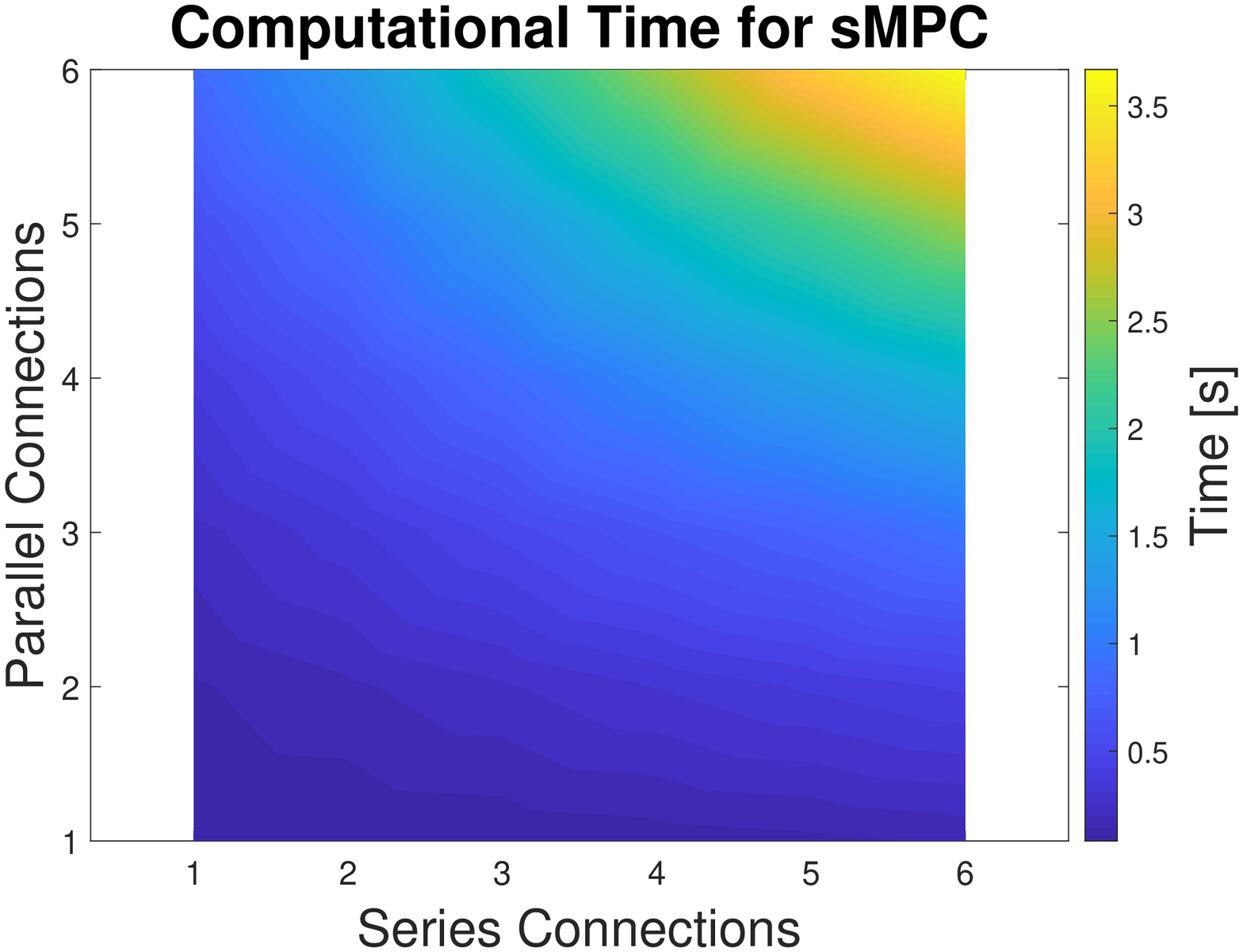}\label{fig:3D_time_sensitivity}}
\subfigure[nMPC]{
\includegraphics[width=0.47\columnwidth]{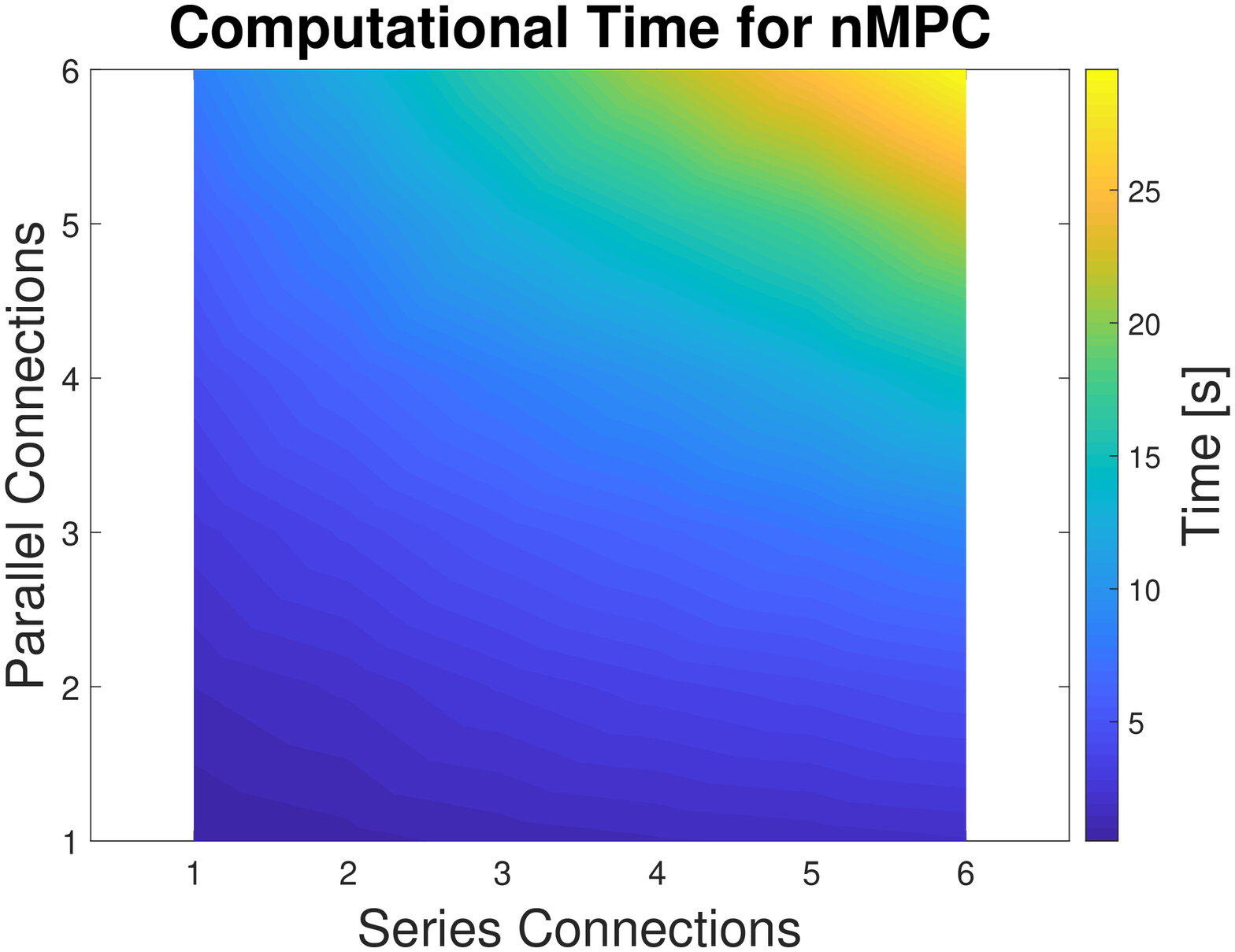}\label{fig:3D_time_nmpc}}
\caption{Mean computational time for the two MPC formulations for up to 6 series and 6 parallel connections. The mean value was computed for hundreds of iterations to ensure statistical significance.}
\label{fig:3D_time_sensitivity_nmpc}
\end{center}
\end{figure}

%
%\begin{figure}[!htb]
%\begin{center}
%\includegraphics[width=\figsize\columnwidth]{fig/mpc/3D_time_sensitivity.eps}
%\caption{Increasing in the computational time of the sensitivity method due to the growth of series and parallel connections.}
%\label{fig:3D_time_sensitivity}
%\end{center}
%\end{figure}
%
%\begin{figure}[!htb]
%\begin{center}
%\includegraphics[width=\figsize\columnwidth]{fig/mpc/3D_time_nmpc.eps}
%\caption{Increasing in the computational time of the nMPC method due to the growth of series and parallel connections.}
%\label{fig:3D_time_nmpc}
%\end{center}
%\end{figure}

\subsubsection{Optimal Control of an Electric Motorbike Battery Pack}
This section demonstrates the applicability of the proposed sensitivity-based MPC to the control of large battery packs, for which nMPC has high on-line computational cost. In this case study, the task is to control the battery of a fully electric motorbike, namely, the electric Vespa Piaggio with a stored energy of $86\,$Ah and a nominal voltage of $48\,$V (i.e., a configuration with 156 cells, arranged in 13 series-connected modules, each with 12 parallel-connected cells, for the Kokam SLPB 75106100). The mean computational time required for each iteration of nMPC was $250\,$s,  which is incompatible with the desired sampling time ($T_s=40\,$s). On the other hand, the mean computational time of sMPC was $30\,$s, which is less than the sampling time.

\section{Conclusions}\label{sec:conclusions}
This work addresses the optimal charging of a battery pack composed of several cells arranged in series and parallel connections. Each cell is described through an electrochemical model that includes kinetics, mass transport, and thermal effects in order to capture the internal physicochemical phenomena. A nonlinear model predictive control (MPC) formulation is formulated that achieves high performance while ensuring constraint satisfaction. An alternative sensitivity-based MPC formulation is proposed that has very similar closed-loop performance but greatly reduces the online computational cost, which makes optimal model-based control suitable for a real-time implementation on a battery pack composed of dozens of cells. The effectiveness of the strategy is demonstrated. The sensitivity-based MPC is successful in providing real-time optimal charging for a fully electric motorbike composed by 156 cells.

%\section{Acknowledgement}\label{sec:acknowledgement}
%Grateful acknowledgement is made to Massimo Zambelli and Prof. Antonella Ferrara (University of Pavia)  for their valuable suggestions and for their contribution in the development of the model used in the proposed control algorithm.

\bibliography{sensitivity_bib}

\end{document}